\documentstyle[11pt,aaspp4]{article}  

\begin{document}

\title{EARLY EVOLUTION OF THE GALACTIC HALO REVEALED
       FROM HIPPARCOS OBSERVATIONS OF METAL-POOR STARS}

\author{Masashi Chiba}
\affil{National Astronomical Observatory, Mitaka, Tokyo 181, Japan}

\and

\author{Yuzuru Yoshii\altaffilmark{1}}
\affil{Institute of Astronomy, Faculty of Science, University of Tokyo, 
Mitaka, Tokyo 181, Japan}
\altaffiltext{1}{Also at Research Center for the Early Universe,
Faculty of Science, University of Tokyo, Bunkyo-ku, Tokyo 113, Japan}

\begin{abstract}
The kinematics of 122 red giants and 124 RR Lyrae variables in the solar 
neighborhood is studied using accurate measurements of their proper
motions by the Hipparcos astrometry satellite, combined with the published
photometric distances, metal abundances and radial velocities.
A majority of these sample stars have metal abundances with [Fe/H]$\le -1$
and thus represent the old stellar populations in the Galaxy.  The halo 
component with [Fe/H]$\le -1.6$ is characterized by no systemic rotation 
$(<U>, <V>, <W>)=(16\pm 18, -217\pm 21, -10\pm 12)$ km s$^{-1}$ and a 
radially elongated velocity ellipsoid $(\sigma_U, \sigma_V, \sigma_W)= 
(161\pm 10, 115\pm 7, 108\pm 7)$ km s$^{-1}$.  About 16\% of such metal-poor 
stars have low orbital eccentricities $e<0.4$, and we see no evidence 
for the correlation between [Fe/H] and $e$.
Based on the model for the $e$ distribution of orbits, we show that
this fraction of low $e$ stars for [Fe/H]$\le -1.6$ is explained from
the halo component alone, without introducing
the extra disk component claimed by recent workers.
This is also supported by no significant change of the $e$ distribution
with the height from the Galactic plane.
In the intermediate metallicity range $-1.6<$[Fe/H]$\le -1$,
we find only modest effects of stars with disk-like kinematics on both
distributions of rotational velocities and $e$ for the sample at $|z|<1$ kpc.
This disk component appears to comprise only
$\sim 10$\% for $-1.6<$[Fe/H]$\le -1$ and
$\sim 20$\% for $-1.4<$[Fe/H]$\le -1$.
It is also verified that this metal-weak disk has the mean rotation of
$\sim 195$ km s$^{-1}$ and the vertical extent of $\sim 1$ kpc,
which is consistent with the thick disk dominating at [Fe/H]$=-0.6$ to $-1$.
We find no metallicity gradient in the halo, whereas there is an indication
of metallicity gradient in the metal-weak tail of the thick disk.
The implications of these results for the early evolution of the Galaxy
are also presented.

\end{abstract}


\section{INTRODUCTION}

Our understanding of how disk galaxies, like our own, were formed has 
greatly advanced in recent years.  Modern large telescopes armed with 
sensitive detectors are about to reach epochs of galaxy formation.  
Ultra faint imagings in the deep Universe have revealed a number of blue, 
irregularly-shaped disks, occasionally accompanying fuzzy blobs 
(Williams et al. 1996).  
Follow-up spectroscopic studies have confirmed that these disk-like systems 
are indeed rotationally supported (e.g. Vogt et al. 1996).
Another line of evidence for forming disk galaxies has emerged from the 
studies of quasar absorption line systems (Pettini et al. 1995;
Lu et al. 1996).  
These absorbers associate heavy elements with abundances much less than 
the solar abundance, thereby implying that we might be seeing the early 
stage of galaxy formation (Lanzetta et al. 1995).
Thus, these deep surveys of high-redshift objects will provide new insight 
into how disk galaxies were formed and how they have evolved to what we see 
today.

Besides in the deep realm of the Universe, our own Galaxy offers more direct
information on the dynamical processes leading to the formation of disks and 
halos in galaxies.  The space motions of old stellar populations observed at 
the current epoch retain the fossil records of the dynamical state in the
early Galaxy, because the relaxation time of the stars exceeds the age of
the Galaxy.  Since formation history of these old stars is imprinted in
their metal abundances, it is possible to know how the Galaxy
has structured while changing the dynamical state with time.

This avenue of research was pioneered by Eggen, Lynden-Bell and Sandage 
(1962, hereafter referred to as ELS).  In their sample consisting of 
nearby disk and high-velocity stars, ELS found the close relationship 
between orbital motions and metallicities in the sense that more metal-poor 
stars have larger orbital eccentricities.  This result led them to conclude 
that the Galaxy collapsed in a free-fall time ($\sim2\times10^8$ yr).  
Various subsequent workers assembled more stellar data based on unbiased
sampling and analyzed the data more rigorously (e.g., Yoshii \& Saio 1979; 
Norris, Bessell, \& Pickles 1985; Norris 1986; Sandage \& Fouts 1987; 
Carney, Latham \& Laird 1990; Norris \& Ryan 1991; Beers \& Sommer-Larsen 
1995).  Then an alternative picture has emerged that the collapse of the 
Galaxy occurred only slowly, lasting much longer than a free-fall time, 
say $\sim 10^9$ yr.  This picture is also supported by a large spread of 
a few $10^9$ years in the ages of both globular clusters and field halo 
stars (Searle \& Zinn 1978, hereafter SZ; Schuster \& Nissen 1989).  
SZ have especially argued that the Galactic halo was not formed in an 
ordered collapse but from the merger or accretion of numerous fragments 
like dwarf-type galaxies. 

It has also been made clear that the Galaxy has an intermediate rapidly
rotating disk or the thick disk having a vertical scale height of 
$\sim 1$ kpc compared to $\sim 350$ pc for the old thin disk (Yoshii 1982;
Yoshii et al 1987; Gilmore \& Reid 1983).  The thick disk is usually 
considered to dominate stars in a range from [Fe/H]$=-0.6$ to $-1$ 
(Freeman 1987), but whether it has a significant metal-weak tail down 
to [Fe/H]$=-2.0$ is a current topic related to this extra disk component 
(e.g., Morrison, Flynn, \& Freeman 1990, hereafter MFF; Beers \& 
Sommer-Larsen 1995).

These kinematical approaches require, among others, the reliable data of 
three dimensional positions and velocities of stars.  In an effort to 
diminish any systematic errors in these basic quantities, the Hipparcos 
satellite was launched in 1989 for the purpose of obtaining accurate 
trigonometric parallaxes and proper motions for numerous bright stars 
distributed over the whole sky.  The Hipparcos satellite is characterized 
by its high accuracy in astrometric measurements to a level of $\sim 1$ 
milliarcsec (mas) for parallaxes and $\sim 1$ mas/yr for proper motions 
(ESA 1997).

Here we revisit the kinematics of red giants and RR Lyrae stars in the solar
neighborhood. The astrometric observations of these stars have been parts of
the Hipparcos programme assigned to the senior author's proposals submitted 
in 1982.  A majority of stars in the sample are characterized by their low 
metallicities with [Fe/H]$<-1$, and are thus thought to represent the 
old halo population in the Galaxy.  Although this sample constitutes only 
a small subset of whole halo stars, great advantage of using it is offered by
the highest accuracy ever achieved in the data of proper motions measured by
Hipparcos. Therefore, combined with a number of well-calibrated photometric
and spectroscopic determinations of 
metal abundances, radial velocities, and distances, this sample may allow us 
to elucidate a more precise picture on the early evolution of the Galactic 
halo.

In Sec. 2, we describe the selection of our sample stars for the Hipparcos 
observations, together with other available data such as metal abundances and 
radial velocities.   The qualities of the obtained astrometric data are 
examined and the effects of the accurate astrometric observations on the 
resulting kinematics of stars are discussed.   Sec. 3 is devoted to the 
kinematical properties of the sample stars and it is explored whether there 
is a signature of the metal-weak thick disk that has recently been discussed.
Sec. 4 is devoted to the orbital motions of the sample stars using the model
gravitational potential of the Galaxy.  We present the distribution of 
orbital eccentricities as a function of metallicity and use it as a 
tool of discriminating the halo from the metal-weak tail of the thick disk.
In Sec. 5 we examine whether a large-scale metallicity gradient exists in 
the Galaxy.   The results of the present paper are summarized and their 
implications for the formation and evolution of the Galaxy are discussed 
in Sec. 6.

\section{OBSERVATIONAL DATA}
\subsection{Star selection}

Red giants used in this paper have been selected from the kinematically 
unbiased sample of metal-deficient red giants surveyed by Bond (1980), 
and RR Lyraes from the catalogues of variable stars compiled by Kukarkin 
(1969-1976).  The sample stars, originally containing 125 red giants and 
362 RR Lyraes, were proposed for the observations with the Hipparcos 
astrometry satellite by one of the authors in 1982.

The sample of red giants consists of stars having apparent $V$ magnitudes 
brighter than $m_V$=12 mag and metal abundances lower than [Fe/H]$=-1.5$. 
The Bond survey is essentially complete to this magnitude, although the 
metal abundances of some of these stars have been significantly revised 
in subsequent studies, as described later.  The sample of RR Lyraes 
consists of almost all stars with $m_V \le$12.5 mag in the Kukarkin 
catalogues.
Only 173 out of 362 RR Lyraes while all of 125 red giants were actually
observed with the Hipparcos satellite.  

In order to analyze the three-dimensional motions of these stars as 
a function of metal abundance, the data of photometric distances, 
radial velocities and metal abundances have 
additionally been assembled from a number of published works.  
At the time of this writing, a complete set of such data is available 
for 122 red giants and 124 RR Lyraes in our Hipparcos sample.

Combining these available data with the Hipparcos measurements of 
parallaxes and proper motions, we made a complete date set as tabulated
in Table 1.   The Hipparcos numbers and the common names of our program 
stars are given in columns 1 and 2, respectively.  The observed values of 
various quantities together with their standard $1\sigma$ errors are 
tabulated in columns 3 to 10.  The literature code numbers are given 
in column 11 for `DA'(photometric distances and metal abundances), 
`V' (radial velocities), and `P' (ground-based proper motions), and 
their correspondence is summarized in Table 2. 

\subsection{Parallaxes and proper motions}

The trigonometric parallaxes $\pi$ and proper motion components
($\mu_{\alpha^\ast}=\mu_\alpha \cos\delta,\ \mu_\delta$) have been measured 
at the catalogue epoch J1991.25 with the Hipparcos satellite, and their
values for our program stars are in columns 3 to 5 of Table 1, together
with the errors which are typically $\sim$1 mas for $\pi$ and $\sim$1 mas/yr
for $\mu$.  

We note that a majority of the stars are located beyond 100 pc from the Sun 
and the relative errors of $\sigma_\pi/\pi$ are larger than 10\% in the 
Hipparcos measurements of parallaxes.  In particular, for much distant 
stars for which true parallaxes are much smaller than their errors, negative 
values have been assigned to the observed parallaxes.  In order to see the 
systematic errors relevant to the Hipparcos observations of our program 
stars, we show in Fig. 1 the relation between $\sigma_\pi/\pi$ and 
$\pi$ (mas) for red giants (filled circles) and RR Lyraes (open circles).
It appears that $\log \sigma_\pi/\pi$ decreases linearly with $\log \pi$,
and this relation virtually agrees with that found for more than 107000 
stars acquired from the first 30 months' observations with Hipparcos 
(Perryman et al. 1995).  Thus the large errors of $\sigma_\pi/\pi$ for our 
sample stars are consistent with the general trend of the Hipparcos accuracy, 
not suffering from some peculiarities inherent in red giants or RR Lyraes.
Nevertheless in order to take advantage of the Hipparcos measurements of
parallaxes, we adopt the direct determination of distances for our program 
stars provided that the relative errors of $\sigma_\pi/\pi$ are less than
20\%.  This condition is fulfilled for only five red giants (HIC\#5445, 
5458, 29992, 68594, 92167) and one RR Lyrae star (HIC\#95497).  For other 
stars, we use the photometric distances.

Prior to the launch of the Hipparcos satellite, various ground-based 
observations had measured proper motions for many of our program stars. 
These proper motions are taken from those listed in the Hipparcos Input 
Catalog (Turon et al. 1992).  If not listed, they are otherwise taken 
from the recently completed catalog of the Lick Northern Proper Motion 
(NPM) program, the NPM1 Catalog (Klemora, Jones \& Hanson 1993) where 
the measurements are accurate to $\sim 5$ mas/yr on the average.  We list 
these ground-based proper motions, if available, in columns 9 and 10 of 
Table 1. 

To examine the quality of the Hipparcos data, we show in Fig. 2 the 
difference between the previous and the Hipparcos measurements for proper 
motion components.  Filled and open symbols represent red giants and 
RR Lyraes, respectively.  For both cases, circles show the stars of
small errors in proper motions 
($|\mu_{\alpha^\ast}|>\sigma_{\mu_{\alpha^\ast}}$ and 
 $|\mu_{\delta}|>\sigma_{\mu_{\delta}})$, and triangles show those
of large errors
($|\mu_{\alpha^\ast}| \le \sigma_{\mu_{\alpha^\ast}}$ or 
 $|\mu_{\delta}| \le \sigma_{\mu_{\delta}}$).
While previous measurements of $\mu_{\alpha^\ast}$ and $\mu_{\delta}$ 
for stars having large proper motions are compatible with the Hipparcos 
measurements, we see large, systematic difference between the previous
and the Hipparcos measurements for stars with small proper motions.
This indicates that the new Hipparcos measurements of higher accuracy
will give insight into the kinematics of stars with small proper motions. 
We note that these stars, containing those with non-eccentric orbits, 
are of particular importance to clarify the formation process of the Galaxy.

\subsection{Distances and abundances}

\subsubsection{Red giants}

The absolute $V$-magnitudes $M_V$, photometric distances $D$, and metal 
abundances [Fe/H] of our red giants have been derived by Bond (1980) based on
the Str\"omgren $uvby$ photometry.  Corrections for the Galactic reddening 
were estimated from a simple csc $|b|$ model, where $b$ is the Galactic 
latitude of the star.  Some stars of the original Bond's (1980) sample have 
been reanalyzed by Carney \& Latham (1986) using the same procedure as Bond
(1980) used, and by Norris, Bessell and Pickles (1986, hereafter NBP) using 
the DDO photometry.  Recently, Anthony-Twarog and Twarog (1994, hereafter 
ATT) largely updated the values of $D$ and [Fe/H] for most stars in the 
Bond's (1980) sample.  ATT obtained new $uvby$ photometries with use of CCDs 
and estimated the realistic reddening effects on red giants using the maps of 
Burstein \& Heiles (1982).  The revised photometric metal abundances appear 
to be in excellent agreement with those of high-dispersion spectroscopy.
It was also pointed out that the metallicity calibration of the DDO 
photometry by NBP and MFF provides reliable [Fe/H] estimates only near 
$-0.8$ and $-2.3$, but systematically underestimates the metallicity 
by about 0.5 dex at [Fe/H]$_{DDO}\sim-1.2$ (Twarog \& Anthony-Twarog 1994, 
1996; Ryan \& Lambert 1995).  This point raises an important issue on the 
existence of metal-poor stars with disk-like kinematics as discussed in 
Sec. 3.2.

For our red giants, we adopt ATT's estimates of $D$ and [Fe/H] except for 
four stars (HIC\# 5458, 38621, 65852, 71087) which ATT did not analyze. 
We adopt Bond's (1980) estimates for such stars.  A standard error in 
[Fe/H] is assigned to be 0.16 dex, which is a typical difference between 
the photometric and the spectroscopic abundances in the ATT sample.  For 
some of our red giants, we use the spectroscopic abundances and associated
errors which have been determined by previous workers and compiled by ATT.
A standard relative error in the derived distances is assigned to be 0.08. 
This is because the ATT calibration of $M_V$ is based on the color $B-V$  
to absolute magnitude $M_V$ relation in the work of NBP where $M_V$ has
a typical error of 0.4 mag. 

We here compare the photometric distances with those derived from the 
Hipparcos parallaxes, using five red giants (HIC\#5445, 5458, 29992, 68594, 
92167) for which the Hipparcos parallaxes are relatively small 
($|\sigma_\pi/\pi|\le 0.2$).   The mean difference between these distances 
is found to be only 15 pc with a dispersion of 55 pc, giving a 25-26\% 
relative error in the distances.  This level of uncertainty may be acceptable 
if a typical error of 8\% in their photometric distances is also taken into 
account.  On the contrary, we necessarily use the photometric distances for 
the stars with larger parallax errors because we see no correlation between 
their photometric and parallactic distances. 

\subsubsection{RR Lyraes}

The metal abundances [Fe/H] of our RR Lyraes are taken from the work of 
Layden (1994).  These values have been measured from the strength of the 
Ca~II~K line relative to the Balmer lines after calibrating it 
to the [Fe/H] abundance scale for the globular clusters studied by Zinn \& 
West (1984).  A typical error in [Fe/H] is $0.15-0.2$ dex.  For some of our 
RR Lyraes which were not observed by Layden (1994), we adopt the [Fe/H] 
values which were estimated by Layden(1996) using the published $\Delta S$ 
values.  The intensity-mean apparent $V$ magnitude and interstellar 
reddenings are taken from the work of Layden (1996) based on the Clube \& 
Dawe (1980) photometry and the reddening maps of Burstein \& Heiles (1982), 
Blanco(1992) and FitzGerald (1968,1987).  

To determine the photometric distances $D$ to our RR Lyraes, we calibrate 
their absolute $M_V$ magnitudes with [Fe/H] assuming a linear relation 
$M_V = a {\rm [Fe/H]} + b$, where the slope $a$ and the intercept $b$ are 
both constants.  There have been many approaches to determine $a$ and $b$, 
including Baade-Wesselink analyses, main sequence fitting of globular 
clusters, and statistical parallax method (for details see, e.g., Carney, 
Storm \& Jones 1992, hereafter CSJ; Layden 1996).  It is seen from Fig. 7 
of Layden (1996) that various $M_V$-[Fe/H] relations lie between the 
relation by CSJ ($a=0.15, b=1.01$) giving the faintest $M_V$ and that of 
Sandage (1993) ($a=0.30, b=0.94$) giving the brightest $M_V$.  
The typical magnitude difference between these two extrema changes from 
$\Delta M_V\approx 0.15$ to 0.37 mag when [Fe/H] decreases from $-0.5$ to 
$-2.0$ dex.  We simply take a mean of these extreme $M_V$ values, because the 
present analysis is not very sensitive to whichever $M_V$-[Fe/H] relation is 
adopted.  The difference between this mean and either of two extreme $M_V$ 
values, which dominates over the error originated from the measurement 
error in [Fe/H], is used as a standard error in $M_V$.

We note that the errors in $M_V$ are the main source of uncertainties in 
estimates of photometric distances.  The relative errors of these distances 
are turned out to be only less than 10\% and are more accurate compared even 
with those derived from the Hipparcos parallaxes (see Fig. 1).  Therefore, 
we use the photometric distances for our RR Lyraes, except for HIC\#95497 
which was observed most accurately with the Hipparcos satellite.  
The small parallax error $|\sigma_\pi/\pi|\le 0.14$ of this star amounts to 
only a 6.9-7.5\% relative error in the distance.  Similarly to red giants,
other RR Lyraes having much larger parallax errors show no correlation 
between their photometric and parallactic distances.

The distributions of distances and metal abundances of our program stars
are shown in panels $a$ and $b$ of Fig. 3, respectively, where the shaded
histogram is for red giants and the open histogram for RR Lyraes.
It is apparent that the stars are sampled mostly within $\sim2$ kpc from 
the Sun.   The metal abundances of red giants are less than [Fe/H]=$-1$ with 
a mean of $-1.8$, whereas those for RR Lyaes are peaked at [Fe/H]$\sim-1.5$
showing a long tail on both sides of the peak metallicity.
It should be noted that metal abundances of red giants extend above the 
limit [Fe/H]$=-1.5$ in the original Bond's (1980) analysis, owing to our
use of the revised metallicity calibration by ATT.

The metallicity distribution for a much larger sample of field halo stars 
was derived by Laird et al. (1988) and Ryan \& Norris (1991),
which involves a small contribution from both old thin disk and
thick disk stars with [Fe/H]$>-1$ (solid line in Fig.3$b$).
Such a distribution is not dissimilar to
that for our whole program stars of red giants and RR Lyraes.
It is therefore suggested that in the metallicity range of [Fe/H]$<-1$
possible incompleteness in our sample may not affect
the following analysis\footnote{
This statement is valid only if there is no age difference between the
halo and thick disk, which affects the RR Lyrae contributions in a
metallicity range relevant to the thick disk. In this respect, in the later
part of the paper, we obtain almost the same contribution of red giants and
RR Lyraes to the metal-weak thick disk at $-1.6<$[Fe/H]$\le-1$ (see Table 5).
This may support no significant age difference, at least for [Fe/H]$<-1$.}.

\subsection{Radial velocities}

A number of previous workers measured radial velocities $V_{rad}$
for our red giants with different accuracies. These include Bond (1980), 
NBP, Carney \& Latham (1986), Barbier-Brossat (1989), and others, as listed 
in Table 2.  If only one work reports $V_{rad}$ for a certain star, 
we simply use it together with the published value of $\sigma_{V_{rad}}$, 
or with $\sigma_{V_{rad}}=5$km s$^{-1}$ if not given.  If more than one 
work report $V_{rad}$'s for a certain star, we adopt the value of 
$V_{rad}$ having the smallest $\sigma_{V_{rad}}$ if given.  Otherwise we 
take the mean of $V_{rad}$'s and estimate $\sigma_{V_{rad}}$ from the 
standard dispersion from the mean.   In the latter case, more than one 
code numbers are listed for literature sources in column 11 of Table 1.
For our RR Lyraes, however, the primary source of $V_{rad}$ and 
$\sigma_{V_{rad}}$ is the work of Layden (1994).

\section{KINEMATICS}
\subsection{Individual and systematic motions}

Given a set of distance, proper motion, and radial velocity of each star, 
we derive its three-dimensional space velocity components $U$, $V$, and 
$W$ directed to the Galactic anticenter, the rotation, and the north pole,
respectively.  These velocity components are corrected for the local solar
motion $(U_{\odot},V_{\odot},W_{\odot})=(-9, 12, 7)$ km s$^{-1}$ with 
respect to the Local Standard of Rest (LSR) (Mihalas \& Binney 1981). 
Associated errors in $(U,V,W)$ are calculated using the formulation of 
Johnson \& Soderblom (1987).  We also derive the velocity components 
$(V_R,V_\phi)$ and their errors in the cylindrical rest frame $(R,\phi)$, 
under the assumption that the solar distance away from the Galactic 
center is $R_{\odot}=8.5$ kpc and the rotational speed of the LSR is 
$V_{LSR}=220$ km s$^{-1}$.

Figure 4 shows the $U$, $V$, and $W$ velocities of the individual stars 
as a function of [Fe/H].  We note that our RR Lyrae sample largely overlaps 
with Layden (1995)'s sample, and the velocity distribution shown here looks 
similar to what is displayed in his paper.  This suggests that the kinematics 
of RR Lyraes based on the previous proper-motion surveys (NPM1; Wan, Mao \& Ji
1980) may remain unchanged even using the Hipparcos proper motions.

It is evident from this figure that metal-poor stars with [Fe/H]$<-1$ have 
large random motions compared to stars with [Fe/H]$>-1$, thus indicating
that the kinematical properties change rather abruptly at [Fe/H]$\sim -1.2$ 
to $-1$, as claimed by Yoshii \& Saio (1987) and subsequently confirmed by 
MFF from their sample of red giants and Layden (1995) from his sample of
RR Lyraes.  This may suggest that the formation of disk component with 
[Fe/H]$>-1$ was distinct from that of the more metal-poor halo component.  

Table 3 shows the mean velocities $(<U>, <V>, <W>)$ and velocity dispersions
$(\sigma_U, \sigma_V, \sigma_W)$ of stars in various metallicity ranges.
The velocity dispersion is estimated from the standard deviation from the
mean after corrected for the observational errors.  It is evident that more 
metal-poor stars are characterized by larger $|<V>|$, that is, larger 
rotation lag behind the LSR, and larger velocity dispersions.  In particular 
metal-poor stars with [Fe/H]$\le -1.6$, which may well represent the halo 
component, have no net rotation ($V_{LSR}-|<V>|\approx 3\pm 21$ km s$^{-1}$) 
and no systematic motions in other velocity components within a range of 
errors ($<U>=16\pm 18$ km s$^{-1}$, $<W>=-10\pm 12$ km s$^{-1}$).
The velocity ellipsoid for these stars is radially elongated giving 
$(\sigma_U,\sigma_V,\sigma_W)=(161\pm 10,115\pm 7,108\pm 7)$ km s$^{-1}$
in reasonable agreement with previous results (e.g., Beers \& Sommer-Larsen 
1995).  The shape of the velocity ellipsoid appears unchanged even if we 
adopt a more restricted metallicity criterion of either [Fe/H]$\le -1.8$ or 
$-2$ for selecting the halo stars.  We note that the $V_R$ and $V_\phi$ 
velocity components in the cylindrical rest frame are essentially the same 
as $U$ and $V_{LSR}+V$, respectively, because our program
stars are localized in the solar neighborhood (Fig. 3$a$).

To examine more closely the rotational properties, we plot $<V_\phi>$, 
$\sigma_\phi$, and $<V_\phi>/\sigma_\phi$ against [Fe/H] in Fig. 5, 
and their values are tabulated in Table 4.  Here the ratio 
$<V_\phi>/\sigma_\phi$ measures how much the system is rotationally 
supported.  It is clearly seen that the rotational properties change rather 
discontinuously at [Fe/H]$\sim -1.4$ to $-1$.  For [Fe/H]$>-1$ there is an 
indication that $<V_\phi>$ correlates with [Fe/H], although the small number 
of these metal-rich stars ($N=17$) makes its significance less definite.
On the other hand, for [Fe/H]$<-1.4$ there is no obvious variation in 
$<V_\phi>$ and $<V_\phi>/\sigma_\phi$ with decreasing [Fe/H], and $<V_\phi>$ 
is consistent with zero rotation within $1\sigma$ errors.  This result
confirms the earlier conclusions (Norris 1986; Carney 1988; Zinn 1988;
Norris \& Ryan 1989) that invalidate the Sandage and Fouts (1987) result
of a linear dependence of $<V_\phi>$ on [Fe/H]. Thus the ELS
hypothesis of a monolithic free-fall collapse from the halo to disk is
not supported (see Norris \& Ryan 1989 for detailed discussion).
It is interesting to note that the stars with $-1.4\le$[Fe/H]$\le-1$ have a
slightly larger $<V_\phi>$ than for the more metal-poor stars, as also 
realized by Layden (1995) from his sample of RR Lyraes.  In the next
subsection we will investigate in more detail whether this suggests an 
intermediate component between halo and thick disk, or more specifically, 
whether this manifests a metal-weak tail of the thick disk component.

\subsection{Is there a metal-weak thick disk?}

MFF advocated from their sample of red giants that there are a significant
number of stars with disk-like kinematics but with low metallicity in a 
range of $-1.6\le$[Fe/H]$\le -1$ near the Galactic plane $|z|<1$ kpc.  
They found that this ``metal-weak thick disk'' (MWTD) rotates rapidly at 
$V\approx 170$ km s$^{-1}$ accounting for about 72\% of the stars in this 
metallicity range, whereas they found no evidence of the MWTD for RR Lyraes. 
Rodgers and Roberts (1993) also argued for the MWTD from their finding of
a large number of candidate blue-horizontal-branch (BHB) stars with
disk-like kinematics [but see Wilhelm (1995) for his different results
using BHB stars, as discussed in Layden (1995)].
However Layden (1995) found that a modest fraction of his sample of 
RR Lyraes show disk-like kinematics only in the metallicity range of 
$-1.3\le$[Fe/H]$\le -1$.  Beers and Sommer-Larsen (1995) argued that the 
MWTD component was confirmed from their large sample of metal-poor stars.
Their MWTD, rotating at $V\approx 195$ km s$^{-1}$, accounts for about
60\% of the stars in the range of $-1.6\le$[Fe/H]$\le -1$ in the
solar neighborhood, and it possess an extremely metal-weak tail down to
[Fe/H]$\le -2$. However, because of the heterogeneous nature of their sample
that includes various types of stars in different evolutionary phases, it is 
not obvious whether all types of stars or only subsamples like red giants 
comprise a large fraction in the MWTD component.  On the other hand, 
ATT demonstrated that the metal abundances of red giants in the range of
$-1.6\le$[Fe/H]$\le -1$ had been underestimated by at most 0.5 dex in the 
DDO photometry of NBP and MFF.  Thus, many of the stars that were previously 
assigned to the MWTD belong to the more metal-rich old disk and/or thick 
disk with [Fe/H]$>-1$.  This suggests that claimed evidence for the MWTD 
component might be less significant than previously thought (see also the 
further discussion in Ryan \& Lambert 1995; Twarog \& Anthony-Twarog 1996).

In view of these controversies, we examine whether our sample of metal-poor 
stars, especially red giants having updated metal abundances and kinematics, 
supports the existence of the MWTD component.  Following the procedure 
adopted by MFF and later workers, we divide our sample into stars at 
$|z|<$1 kpc and $|z|\ge$1 kpc and show the frequency distribution of 
$V_\phi$ for four metallicity intervals in Fig. 6.  Solid and dashed 
histograms represent red giants and RR Lyraes, respectively. 
At $|z|<$1 kpc the metal-rich stars with [Fe/H]$\ge -1$ (panels $a$ and 
$b$) are characterized by a high rotational velocity of $V_\phi=200$ to 
220 km s$^{-1}$.  Because of the small number of such stars, it is not 
clear whether the $V_\phi$-velocity distribution has a Gausssian nature 
as MFF reported.   For the stars of our concern in the metallicity range of 
$-1.6\le$[Fe/H]$\le -1$ (panel $c$), we are unable to verify MFF's 
finding of the strongly asymmetric $V_\phi$-velocity distribution (their 
Fig. 7$c$) which they considered as an evidence for the MWTD component.  
There is indeed an indication that the $V_\phi$-velocity distribution for 
our red giants is somewhat skewed towards positive $V_\phi$, but is not as 
significant as demonstrating the MWTD component.   Furthermore, the 
$V_\phi$-velocity distribution for our red giants is similar to those 
derived by MFF and Layden (1995) from their samples of RR Lyraes where
a possible contribution of the MWTD was already shown to be modest in such
a metallicity range.  For the more metal-poor stars with [Fe/H]$\le -1.6$ 
(panel $d$), the $V_\phi$-velocity distribution for the composite sample of
red giants and RR Lyraes is essentially the same as that presented in MFF, 
and this is also the case for stars at $|z|\ge$1 kpc (panels $f$ and $g$).

To examine more quantitatively the existence of the MWTD component in 
the stars with $-1.6\le$[Fe/H]$\le -1$ (panel $c$), we fit the data to 
a mixture of two Gaussians representing the separate components of halo 
and disk.  Under the assumption that the mean velocity $<V_\phi>_{halo}$ 
and velocity dispersion $\sigma_{\phi,{halo}}$ for the halo are fixed as 
those for the stars with [Fe/H]$<-1.6$ in panel $d$, we evaluate the 
best-fit values of the disk quantities such as $<V_\phi>_{disk}$ and 
$\sigma_{\phi,{disk}}$ as well as the disk fraction $F$.  The likelihood 
function for stars with $V_\phi^i$ is then given by (MFF)
\begin{equation}
\log f(F,<V_\phi>_{disk},\sigma_{\phi,{disk}}) =
\sum_i \log [ F f_{disk}^i + (1-F) f_{halo}^i] \ ,
\end{equation}
where
\begin{equation}
f_{disk}^i=\frac{1}{\sigma_{\phi,{disk}}\sqrt{2\pi}}
  \exp[ - (V_\phi^i-<V_\phi>_{disk})^2/2\sigma_{\phi,{disk}}^2 ] \\,
\end{equation}
and 
\begin{equation}
f_{halo}^i=\frac{1}{\sigma_{\phi,{halo}}\sqrt{2\pi}}
  \exp[ - (V_\phi^i-<V_\phi>_{halo})^2/2\sigma_{\phi,{halo}}^2 ] \ .
\end{equation}

Before applying the maximum likelihood analysis, we determine the halo
quantities $<V_\phi>_{halo}$ and $\sigma_{\phi,{halo}}$ for each of samples 
with [Fe/H]$<-1.6$ consisting of red giants and RR Lyraes.  Using these halo 
quantities as fixed, we then find the best-fit values of the disk quantities 
in three low-metallicity ranges of $-1.6<$[Fe/H]$\le -1$, 
$-1.5<$[Fe/H]$\le -1$, and $-1.4<$[Fe/H]$\le -1$.
Figure 7 shows the $V_\phi$-velocity distribution in these low-metallicity
ranges for red giants (left panels), RR Lyraes (middle panels), and both
stars (right panels).  The data are shown by histograms.   The results of
the maximum-likelihood analysis are shown by lines in Fig. 7 and tabulated
in Table 5.  In sharp contrast with the results by MFF and Beers and 
Sommer-Larsen (1995), we found only a modest disk fraction of $F\sim 0.3$ 
for either red giants or RR Lyraes or both.  It is also interesting to note 
that the derived mean velocity $<V_\phi>_{disk}\approx 118$ km s$^{-1}$ is 
much smaller than previously reported.  

We here attempt to determine the fraction of more rapidly-rotating disk 
at $<V_\phi>_{disk}\approx 195$ km s$^{-1}$ which was postulated as the 
MWTD by Beers and Sommer-Larsen (1995).  For this purpose we further fix 
$<V_\phi>_{disk}=195$ km s$^{-1}$ and find the best-fit values of
$\sigma_{\phi,{disk}}$ and $F$.  The results tabulated in Table 5 indicate
that the fraction of this rapidly rotating disk is only $0.1-0.2$ and 
therefore its existence is quite marginal.  Given such a small fraction of 
the rapidly rotating disk, however, the present analysis alone can not tell 
which type of the MWTD, rotating slowly at 
$<V_\phi>_{disk}\sim 120$ km s$^{-1}$ or rapidly at 
$<V_\phi>_{disk}\sim 200$ km s$^{-1}$, is actually preferred.
We will return to this problem in Sec. 4.4 analyzing the orbital motions of 
stars.

\section{ORBITAL PROPERTIES}

Stars observed in the solar neighborhood have traveled from different, often 
much distant, parts within the Galaxy.  In this section we investigate the 
orbital motions of our program stars in a model gravitational potential of
the Galaxy.  We will especially focus on the distribution of orbital 
eccentricity and use it as diagnostic for studying the global dynamics of 
the Galaxy.

\subsection{Gravitational potential}

We investigate space motions of our program stars in two representative 
types of the gravitational potential which are both axisymmetric and 
stationary.  One is the two-dimensional potential $\Phi(R)_{ELS}$ adopted 
first by ELS and subsequently by most of previous workers.  Although this 
potential gives the projected orbits onto the Galactic plane, we can compare 
the planar orbits of our program stars directly with those previously 
reported.  Another is the more realistic three-dimensional potential that 
allows vertical motion above and below the Galactic plane.  Sommer-Larsen 
and Zhen (1990, hereafter SLZ) adjusted the parameters in the analytic 
St\"ackel potential and reproduced the mass model of Bahcall, Schmidt, \& 
Soneira (1982).  We adopt this potential $\Phi(R,z)_{SLZ}$ because the 
analytic potential has a great advantage for keeping the clarity in the
analysis.

Some cautions are in order for use of $\Phi(R)_{ELS}$.  This potential is 
motivated to reproduce the mass distribution in the disk without including 
a massive dark halo.  Thus, some stars with large velocities or in highly 
eccentric orbits become unbound in the original ELS potential.  In order to 
effectively take into account the effects of a massive halo, we derive the 
escape velocity $V_{esc}$ from our sample stars and put a new constraint on 
$\Phi(R)_{ELS}$.

Three red giants are found to possess the rest-frame velocity in 
excess of 400 km s$^{-1}$, that is, HIC69470 (437 km s$^{-1}$), HIC75263 
(454 km s$^{-1}$), and HIC104191 (562 km s$^{-1}$).   For the extremely 
high-velocity star HIC104191 (HD200654), however, there appears a large 
discrepancy among the estimates of $D$ and [Fe/H] by ATT, NBP, and Bond
(1980).  Instead of our use of ATT ($D$,[Fe/H])=(0.463 kpc,$-2.79$), the 
estimates by NBP (0.404 kpc,$-2.26$) and Bond (1980) (0.320 kpc,$-2.40$) 
give the rest-frame velocity of 472 km s$^{-1}$ and 348 km s$^{-1}$, 
respectively.   Since the reason for such a large discrepancy is not 
known, we exclude the star HIC104191 and adopt $V_{esc}=450$km s$^{-1}$ 
in agreement with the result by Sandage and Fouts (1987).  This value of 
$V_{esc}$ is used to constrain $\Phi_{ELS}$ in a way described in Appendix.  
We note that the inclusion/exclusion of this star hardly affects the 
following analysis.

The SLZ potential $\Phi(R,z)_{SLZ}$ consists of two components corresponding
to a flattened perfect oblate disk and a slightly oblate massive halo.  The 
latter is modeled by the analytic $s=2$ model of de Zeeuw, Peletier and Franx
(1986) which gives a density profile $\rho(R=0,z)\propto 1/(z^2+c^2)$ along 
the $z$-axis where $c$ is a constant.  This potential provides a nearly flat 
rotation curve beyond $R=4$ kpc and well reproduces the local mass density 
at $R_\odot$.  We adopt the values of the parameters in $\Phi(R,z)_{SLZ}$
which were determined by SLZ.  We note that the large escape velocity 
$V_{esc}$ reported above can be attributed to the massive halo.  The actual 
value of $V_{esc}$ is reproduced by setting an arbitrary boundary or
tidal radius at the edge of the halo. This method of tuning the potential
gives essentially no quantitative change for the orbital properties of stars
inside the boundary.

\subsection{Eccentricity versus metallicity}

Using a model gravitational potential we compute the orbital eccentricity 
defined as $e = (r_{ap}-r_{pr})/(r_{ap}+r_{pr})$ where $r_{ap}$ and $r_{pr}$ 
denote the apogalactic and perigalactic distances, respectively.  In Fig. 8 
we plot our sample stars in the $e-$[Fe/H] diagram where the eccentricities 
are based on either $\Phi(R)_{ELS}$ (panel $a$) or $\Phi(R,z)_{SLZ}$ (panel 
$b$).  Contrary to the ELS result, there is no apparent correlation between 
$e$ and [Fe/H] for stars with [Fe/H]$\le -1$, as has been claimed by previous
workers (Yoshii \& Saio 1979; NBP; Carney \& Latham 1986; Carney, Latham \&
Laird 1990; Norris \& Ryan 1991).  The orbital motions of stars 
in this metallicity range are dominated by high-$e$ orbits, but a finite
fraction of stars have small-$e$ orbits, even in the range of 
[Fe/H]$\le-1.6$.

We note that the result for [Fe/H]$\le-1.6$ is almost unchanged by the ATT's 
revised [Fe/H] calibration for metal-poor red giants, because this revised 
calibration is only effective at [Fe/H]$\sim -1.2$.   Thus, we conclude that 
the orbital motions of metal-poor halo stars in the solar neighborhood are
indeed characterized by a diverse distribution of eccentricity.  This is 
more clearly demonstrated by showing the differential distribution $n(e)$ 
in Fig. 9 and the cumulative distribution $N(<e)$ in Fig. 10 for either the 
ELS eccentricity (panel $a$) or the SLZ eccentricity (panel $b$), where the 
solid-line and dotted-line histograms represent the stars with 
[Fe/H]$\le-1.6$ and $-1.6<$[Fe/H]$\le-1$, respectively.   Our sample stars
having [Fe/H]$\le-1.6$ and $e<0.4$ comprise 13\% for the ELS eccentricity
and 16\% for the SLZ eccentricity.

Special attention has been paid to search for metal-poor halo stars with 
small-$e$ orbits, because their existence constrains the dynamical evolution
of the Galaxy (ELS; Yoshii \& Saio 1979).  NBP claimed that 20\% of stars 
with [Fe/H]$<-1$ have $e<0.4$ in their non-kinematically selected sample, 
whereas Carney \& Latham (1986) found $5-8$\% in their sample of red giants 
with [Fe/H]$<-1.5$.  Subsequent workers have further obtained a fraction of 
$e<0.4$ ranging from a few to a few tens \% (Carney, Latham \& Laird 1990;
Norris \& Ryan 1991).  In particular, the fraction of such stars has recently
been discussed for examining whether the MWTD is a significant component in 
the Galaxy (Ryan \& Lambert 1995; Norris 1996).

It is worth noting that there are several effects that change the estimated
fraction of metal-poor stars with low eccentricity.  First, a sample selected 
from high proper motion stars has a significant bias against low eccentricity
(Yoshii \& Saio 1979; Norris 1986).  Second, systematic errors in the [FeH]
calibration affect the number of stars counted in the respective range of 
[Fe/H].  Specifically, the previous analyses using the [Fe/H] calibration by 
NBP or MFF for red giants are subject to this effect (Twarog \& Anthony-Twarog
1994; Ryan \& Lambert 1995).  Third, an estimation of $e$ is {\it not} 
insensitive to the Galactic gravitational potential.  Most prior workers
used the original or modified planer ELS potential to obtain the projected
$e$ onto the Galactic plane, except for Yoshii \& Saio (1979) and Carney et 
al. (1990) who used the vertically extended gravitational potential.  
Yoshii \& Saio (1979) demonstrated that use of the planar ELS potential
overestimates $e$.

We note that there is a freedom of changing the basic parameters even 
in the ELS potential.  These are the radial scale length and amplitude 
of the potential, which are scaled by $R_{\odot}$ and $V_{\odot}$, 
respectively.  In their original paper, ELS adopted $R_{\odot}=10$ kpc 
and $V_{\odot}=250$ km s$^{-1}$, and these values have also been used 
by NBP and subsequent workers.  Carney et al. (1990) adapted the ELS 
potential to the updated values of $R_{\odot}=8$ kpc and 
$V_{\odot}=220$ km s$^{-1}$, whereas we use $R_{\odot}=8.5$ kpc and 
$V_{\odot}=220$ km s$^{-1}$ together with an extra constraint on $V_{esc}$ 
(see Appendix).
Table 6 summarizes how these changes of the parameters affect the fraction
of stars with [Fe/H]$\le -1.6$ and $e<0.4$ in our sample.  We see that the 
potential giving more mass density in the solar neighborhood has the effects 
to bind the stars more tightly and to reduce their apogalactic distances and 
eccentricities, so that the number of stars with low eccentricity is 
increased.  Accordingly we emphasize that {\it the reported fraction of
metal-poor stars with $e<0.4$ is inevitably dependent on what form of the 
Galactic gravitational potential is adopted}.

\subsection{Model eccentricity distribution for halo stars}

Given a fraction of low-metal and low-$e$ stars in our sample, we examine 
whether such fraction is consistent with that expected from the velocity 
distribution of halo stars.

In Sec. 3, we obtained the velocity distribution for metal-poor stars in the 
solar neighborhood.  This is represented by an approximately Gaussian, and 
the velocity ellipsoid is radially elongated with 
$(\sigma_U,\sigma_V,\sigma_W)=(161,115,108)$ km s$^{-1}$ for [Fe/H]$<-1.6$. 
The expected distribution of eccentricity is tightly related to this 
velocity distribution for the elongated orbital motions of stars which 
arrive near the Sun.

To demonstrate this situation graphically, we show in Fig. 11 the so-called 
Bottlinger diagram in the $UV$ plane, where solid lines represent the loci 
of constant eccentricity derived from $\Phi_{ELS}$.  Obviously stars having 
the eccentricity less than $e$ are enclosed within a locus of constant $e$ 
in this diagram.   For nonzero $W$ velocities, such stars are enclosed within
a surface of constant $e$ in the full $UVW$ space.  In this way, for a given 
velocity distribution we obtain the corresponding $e$ distribution which 
depends on the adopted form of gravitational potential.   

We assume that the velocity distribution of halo stars is given by a single 
Gaussian with no net rotation:
\begin{equation}
f(U,V,W) = \frac{1}{(2\pi)^{3/2}\sigma_U\sigma_V\sigma_W}
\exp[ - \frac{U^2}{2\sigma_U^2} - \frac{(V+V_{LSR})^2}{2\sigma_V^2}
      - \frac{W^2}{2\sigma_W^2} ] \ ,
\end{equation}
where $V_{LSR}=220$ km s$^{-1}$.  When the planar potential of $\Phi_{ELS}$ 
is used, we can set $W=0$ in eq. (4), and the cumulative $e$ distribution 
$N(<e)$ is obtained by integrating $f(U,V,W=0)$ over the $UV$-plane within 
a locus of constant $e$.
For the three dimensional potential of $\Phi_{SLZ}$, we perform the Monte 
Carlo simulation by creating an ensemble of stars based on $f(U,V,W)$ and 
estimate $e$ for each star. Here the analytic nature of $\Phi_{SLZ}$ has 
a great advantage of quick estimation of the $e$ distribution for numerous 
simulated stars, whereas the procedure is quite time-consuming for a
non-analytic potential for which numerical integrations of orbits are required.

We consider (A) a radially elongated ellipsoid derived from the stars with 
[Fe/H]$<-1.6$, $(\sigma_U,\sigma_V,\sigma_W)=(161, 115, 108)$ km s$^{-1}$, 
and (B) a tangentially elongated ellipsoid $(115, 161, 108)$ km s$^{-1}$ 
motivated for the purpose of comparison by interchanging $\sigma_U$ and 
$\sigma_V$.  The results for these different velocity ellipsoids are shown 
by bold solid and dashed lines, respectively, in Figs. 9 and 10.

It is remarkable that such a radially elongated velocity ellipsoid gives
the $e$ distribution which agrees well with the observation for 
[Fe/H]$\le-1.6$.  This is still the case if we use the potential with
different values of $V_{esc}$, $R_{\odot}$ and $V_{\odot}$, as shown in 
Table 6.  Some slight differences between the model and observed $e$ 
distributions may have arisen from (1) statistical fluctuation owing to the 
smallness of our sample size, (2) weak dependence of the velocity 
distribution on the space coordinates adopted in the analysis, and 
(3) slight deviation from a pure Gaussian velocity distribution.
Nonetheless, the reasonably good fit of the model curve suggests that
{\it the observed $e$ distribution for [Fe/H]$\le-1.6$ and the fraction of 
small-$e$ orbits ($e<0.4$) are naturally explained from a single Gaussian 
velocity distribution of only the halo component characterized by a radially 
elongated velocity ellipsoid}.  This implies that for explaining the existence
of such low-metal and low-$e$ stars it is {\it no longer} necessary to 
introduce the extra MWTD component which extends down to [Fe/H]$\le -1.6$.  
It is interesting to note that if the velocity distribution is tangentially 
anisotropic, as argued by Sommer-Larsen et al. (1994) from their sample stars
at large galactocentric distances, we would observe the $e$ distribution as 
shown by bold dashed curves in Figs. 9 and 10.

Our sample stars in Figs. 9 and 10 are not restricted to stars with small 
errors $\sigma_e$ in derived eccentricities.  NBP imposed the criterion
$\sigma_e \le 0.1$ but this has been claimed to produce an extra bias against
stars with small proper motions or perhaps low eccentricities (Carney \& 
Latham 1986; Twarog \& Anthony-Twarog 1994).  To see whether this is also the 
case in our Hipparcos sample, we plot $\sigma_e$ versus $e$ (ELS) in Fig. 13. 
It is evident that only the intermediate-$e$ orbits ($e=0.4\sim0.5$) suffer
from large errors of $\sigma_e>0.1$.  The relatively small errors for 
small-$e$ orbits may be attributed to the accurate measurements of small 
proper motions by the Hipparcos satellite (see Fig. 2).  Therefore the 
fraction of small-$e$ orbits is unchanged if we confine ourselves to the 
stars with $\sigma_e \le 0.1$.  This criterion $\sigma_e \le 0.1$ instead 
eliminates quite a number of stars with $e=0.4-0.5$ and therefore reduces 
the observed excess over the predicted $e$ distribution seen in Fig. 9,
which further strengthens the conclusion obtained here.

\subsection{Effects of the metal-weak thick disk on the $e$ distribution}

Figure 10 further indicates that the observed fraction of $e<0.4$ stars in 
the metallicity range of $-1.6<$[Fe/H]$\le -1$ appears to be systematically 
larger than that expected solely from the velocity distribution with 
[Fe/H]$\le -1.6$.  In order to see whether this excess belongs to the MWTD
component, we select the stars at $|z|<1$ kpc as in Sec. 3.2 and derive the 
cumulative $e$ distribution $N(<e)$ based on $\Phi_{SLZ}$.  The results for
$-1.4<$[Fe/H]$\le -1$, $-1.6<$[Fe/H]$\le -1$, and [Fe/H]$\le -1.6$ are shown 
by dashed, dotted, and solid histograms respectively in panel $a$ of Fig. 13.
Bold solid line shows the model $N(<e)$ expected from 
$(\sigma_U,\sigma_V,\sigma_W)=(165, 120, 107)$ km s$^{-1}$ for stars at 
$|z|<1$ kpc with [Fe/H]$<-1.6$.  The model again reproduces the observation
for [Fe/H]$\le -1.6$ reasonably well.  

It is evident that the low-$e$ stars with [Fe/H]$>-1.6$, which may belong 
to the MWTD, indeed occupy a larger fraction beyond the prediction at lower 
$e$.  This observed excess is even larger for $-1.4<$[Fe/H]$\le -1$ than that 
for $-1.6<$[Fe/H]$\le -1$.   This and the following result remain unchanged 
even if we use $\Phi_{ELS}$ instead of $\Phi_{SLZ}$.   Similarly as in 
Sec. 3.2, we here attempt to explain this excess component in terms of 
either (1) the rapidly rotating MWTD at $<V_\phi>_{disk}=195$ km s$^{-1}$
or (2) the slowly rotating MWTD at $<V_\phi>_{disk}=120$ km s$^{-1}$. 
The model calculation is performed using a mixture of two Gaussian velocity 
distributions which consist of the non-rotating halo and the rotating MWTD 
at $<V_\phi>_{disk}$.  For the non-rotating halo we adopt the velocity 
dispersion for stars at $|z|<1$ kpc with [Fe/H]$<-1.6$, while the velocity
distribution for the MWTD is taken from Beers and Sommer-Larsen's (1995) 
result $(\sigma_U,\sigma_V,\sigma_W)=(63,42,38)$ km s$^{-1}$ for their 
thick-disk stars at $|z|<1$ kpc.  The MWTD component is assumed to comprise 
the fraction $F$.

Figure 13 shows the model $N(<e)$ distributions for 
$<V_\phi>_{disk}=195$ km s$^{-1}$ (panel $b$) and 
$<V_\phi>_{disk}=120$ km s$^{-1}$ (panel $c$).   It follows from panel $b$ 
that the rapidly rotating MWTD at $<V_\phi>_{disk}=195$ km s$^{-1}$ explains 
the excess for [Fe/H]$>-1.6$, provided that $F$ is as small as a few tenths
[$F=0.2$ for $-1.4<$[Fe/H]$\le -1$ (bold dashed line) or $F=0.1$ for 
$-1.6<$[Fe/H]$\le -1$ (bold dotted line)]\footnote{
Even if we adopt the cooler halo velocity ellipsoid as obtained by some of
previous workers [e.g.
$(\sigma_U,\sigma_V,\sigma_W)=(130, 100, 90)$ km s$^{-1}$], we see only
a few \% change in the value of $F$.}.
Contrary to the claim by MFF 
and Beers and Sommer-Larsen (1995), there is no evidence supporting much 
higher fraction of this MWTD component [see the model for $F=0.6$ (bold 
dash-dotted)].  On the contrary, the slowly rotating MWTD at
$<V_\phi>_{disk}=120$ km s$^{-1}$ fails to reproduce the excess at lower 
$e$ even if $F$ is increased (panel $c$). It is worth noting that the 
likelihood analysis in Sec. 3.2 using the $V_\phi$ distribution yields  
larger $F$ for more slowly rotating MWTD.  This does not necessarily 
indicate that the MWTD component with slower rotation is preferentially
confirmed, since the $V_\phi$ distribution conveys only the partial 
information on the full three-dimensional orbital motions of stars.

We then turn to the question concerning how far the MWTD extends above or
below the disk plane.  In panel $a$ of Fig. 14 we show the cumulative
distribution $N(<e)$ for stars at $|z|\ge 1$ kpc.  It is of particular
interest that the $e$ distribution for Fe/H]$\le -1.6$ (solid histogram) 
remain essentially {\it unchanged} when stars are selected at high $|z|$.
On the other hand, the fraction of small-$e$ orbits is greatly reduced
for both $-1.6<$[Fe/H]$\le -1$ (dotted histogram) and $-1.4<$[Fe/H]$\le -1$ 
(dashed histogram).   This apparent lack of low-$e$ stars at $|z|\ge 1$ kpc 
is seen from the [Fe/H] versus $e$ diagram in panel $b$ of Fig. 14, where
the stars with $e<0.6$ are absent in the metal-poor range of 
$-1.6<$[Fe/H]$\le -1$ and in the metal-rich range of [Fe/H]$>-0.8$
(the region enclosed by dotted line).

More direct insight into the vertical extent of the MWTD is obtained from 
Fig. 15 where the fraction of stars having $e<0.4$ is shown as a function 
of the limiting height $z_{lim}$ above or below which the stars are located, 
i.e., $|z|\ge z_{lim}$.  This fraction sharply drops at $z_{lim}=0.8-1$ kpc 
for stars with $-1.6<$[Fe/H]$\le -1$ (dotted line) or $-1.4<$[Fe/H]$\le -1$ 
(dashed line), whereas it is kept almost constant for stars with 
[Fe/H]$\le -1.6$ (solid line).  Thus, the rapidly rotating MWTD component, 
which we have identified in the metal-poor range of $-1.6<$[Fe/H]$\le -1$, 
have an vertical extent of $0.8-1$ kpc.  This is virtually consistent with 
current estimates of the thick-disk scale height (e.g., Yoshii, Ishida, \& 
Stobie 1987).  It should also be noted that the halo component which is 
exclusively represented by stars with [Fe/H]$\le -1.6$ shows no significant 
$z_{lim}$-dependence (solid line) and no noticeable contamination from the 
rapidly rotating MWTD.   The low-$e$ fraction for [Fe/H]$\le -1.6$ slightly
increases towards higher $z_{lim}$.  This may be explained from the fact 
that a star locating further out from the solar neighborhood has a smaller 
radial range in its orbital motion when a set of integrals of motion is
given (Yoshii \& Saio 1979).

\section{Metallicity gradient as a clue to formation history}

\subsection{Introduction}

Whether the Galaxy has a global metallicity gradient has been another key 
issue for understanding its early evolution.  ELS used the $|W|$ velocity 
as an indicator of the maximum vertical height $|z_{max}|$ that stars can 
reach, and the ultraviolet excess $\delta (U-B)$ as an indicator of the 
metallicity corresponding to the epoch at which stars were born from the 
gas.  ELS therefore argued that the correlation between $\delta (U-B)$ and 
$|W|$ may have arisen if the more metal-poor, older populations were formed 
at systematically larger heights beyond the disk, in other words, the Galaxy 
may have collapsed from an extended gas sphere to the disk.

Their argument implies that presence or absence of a large-scale metallicity 
gradient depends on the balance of competing time scales among the collapse 
of the Galaxy, the metal enrichment, and the spatial mixing of heavy 
elements in the gas (see also Sandage \& Fouts 1987).  In the free-falling 
proto-Galaxy via dissipation, the gas was progressively confined to smaller 
volumes, while newly formed stars were left over out of this infalling gas.  
This indicates higher metallicity for stars that were formed within smaller 
volumes, thereby causing the metallicity gradient.  On the contrary, if the 
Galaxy was formed in a discontinuous or inhomogeneous manner, e.g., by 
merging of numerous fragments such as dwarf-type galaxies which have their 
own chemical histories (SZ), no global metallicity gradient would be 
observed in any spatial directions. 

Figure 4 has already shown that a monotonous increase of the $|W|$-velocity 
with decreasing [Fe/H] is only detectable at [Fe/H]$>-1.4$ but not apparent 
at [Fe/H]$<-1.4$.   It is important to here caution that the $|W|$-velocity 
alone does not characterize $|z_{max}|$ in a three-dimensional gravitational 
potential.  As we demonstrate graphically in Fig. 16, our sample stars 
observed in the solar neighborhood have traveled through more distant 
regions of the Galaxy.  For instance a star now
at $(R,z)\sim(8.6,-0.5)$ kpc can orbit in 
the accessible area enclosed by solid line, whereas a star
at $(R,z)\sim(8.6,0.0)$ kpc
within the dotted-line area.  Since the orbital $z$-motion is coupled with 
those in the $R$ and $\phi$ directions, $|W|$ is not necessarily related 
with $|z_{max}|$, especially for stars having large $|W|$-velocity or large 
asymmetric drift (see Carney et al. 1990).

We estimate the maximum height $|z_{max}|$ and the apogalactic cylindrical 
distance $R_{max}$ for each star using SLZ's gravitational potential, and 
then examine how the estimates of $|z_{max}|$ and $R_{max}$ for our sample 
stars are related with their metal abundances.
We first divide our sample into four $V_\phi$ bins of $V_\phi \le \infty$, 
170 km s$^{-1}$, 120 km s$^{-1}$, and 70 km s$^{-1}$.  The halo component 
among various populations is extracted simply by selecting stars with small 
azimuthal velocity $V_\phi$ (Sandage \& Fouts 1987; Carney et al. 1990, 
see also Sec. 3).  However it is admittedly more problematic to discriminate
the MWTD component alone.  If we select large-$V_\phi$ stars assuming that 
the MWTD is rapidly rotating, the resultant sample will be considerably
contaminated by the old disk component of metal-rich stars with [Fe/H]$>-0.6$.
To avoid such a sampling bias, we attempt to discriminate the MWTD component 
from the small-$V_\phi$ halo component.  We take advantage of the result
in Sec. 4 that the MWTD stars likely have smaller $e$ compared to the halo
stars and that their vertical distribution is confined within $|z|<1$ kpc. 
Accordingly we impose additional constraints of $e\le 0.6$ and $|z|<1$ kpc
to discriminate the MWTD from the halo component.

\subsection{Results}

Plots of [Fe/H] against $R_{max}$ and similar plots against $|z_{max}|$ 
are shown in Figs. 17 and 18, respectively, for ($a$) $V_\phi \le \infty$ 
or all stars, ($b$) $V_\phi \le 170$ km s$^{-1}$, and ($c$) 
$V_\phi \le 120$ km s$^{-1}$.  Note that the $V_\phi$ criteria used in 
panels $b$ and $c$ have successfully selecting halo stars with [Fe/H]$<-1$. 
The mean metal abundances in five bins of $R_{max}$ and in six bins of 
$|z_{max}|$ are connected by solid lines with estimated 1$\sigma$ errors of 
the means.  These data are tabulated in Tables 7 and 8, where the results 
for $V_\phi \le 70$ km s$^{-1}$ are also tabulated.  These figures and 
tables clearly indicate that stars with $V_\phi\le 170$ or $120$ km s$^{-1}$ 
show no large-scale metallicity gradient in the $R$ and $z$ directions 
within a 1$\sigma$ error level.  This agrees with prior works based on 
different samples of field stars (Saha 1985; Carney et al. 1990) and halo 
globular clusters (Zinn 1985).

Figures 19 and 20 show the results for the MWTD candidate stars.  We find
that the additional constraints of $e\le 0.6$ and $|z|<1$ kpc are effective
for excluding the very metal-poor stars with [Fe/H]$<-1.8$.  
In contrast to the halo component as discussed above, there is an indication 
of a metallicity gradient 
$\Delta [Fe/H]/\Delta R_{max} = -0.03 \sim -0.02$ dex kpc$^{-1}$ 
from $R_{max}=7$ to 18 kpc, and 
$\Delta [Fe/H]/\Delta |z_{max}| =-0.07 \sim -0.05$ dex kpc$^{-1}$ from 
$|z_{max}|=1$ to 8 kpc.  The number of the MWTD candidates may not be large 
enough to give a statistically significant result (see Tables 7 and 8), 
however it is intriguing to note that the obtained metallicity gradient is 
larger than the gradient previously detected from the thick-disk stars with
[Fe/H]$\ge -1$ (e.g., Majewski 1993 for review).  Further studies based on 
the assembly of more sample stars shall be important to elucidate this
discrepancy and clarify the formation process of this controversial 
component.


\section{DISCUSSION AND CONCLUSION}

We have investigated the kinematics of 122 red giants and 124 RR Lyraes, 
which were selected without kinematic bias and were observed by the 
Hipparcos satellite to measure their accurate proper motions. 
The metal abundances of our program stars range from [Fe/H]$=-1$ to $-3$, 
thereby suitable for analyzing the halo component as well as the metal-weak 
tail of the thick disk component below [Fe/H]$=-1$.  We summarize the 
obtained results below and discussed them in the light of the early 
evolution of the Galaxy.

\subsection{Summary of the results}

The present analyses indicate that the solar-neighborhood stars with 
[Fe/H]$\le -1$ mostly have the halo-like kinematics of large velocity 
dispersion and no significant rotation.  The velocity ellipsoid is radially 
elongated yielding $(\sigma_U,\sigma_V,\sigma_W)=(161,115,108)$ km s$^{-1}$ 
in the metal-poor range of [Fe/H]$\le -1.6$.  The rotational properties of 
the system probed by $<V_\phi>$ or $<V_\phi>/\sigma_\phi$ appear to change 
largely at [Fe/H]$=-1.4\sim -1$ (Fig. 5), indicating that the collapse of 
the Galaxy from the halo to the disk took place discontinuously.

We have found no correlation between [Fe/H] and $e$ for [Fe/H]$\le -1$ 
(Fig. 8), which is in contrast to the ELS result.   Even for 
[Fe/H]$\le -1.6$, about 16\% of our program stars are found to have $e<0.4$ 
(the value of $e$ slightly depends on the gravitational potential adopted), 
and this fraction of low-$e$ stars stems from the radially elongated 
velocity ellipsoid of the halo component alone, without introducing an 
extra disk component (Figs. 9 and 10).  Thus, the existence of low-$e$ 
stars does not necessarily mean the dominance of an extra rapidly-rotating 
component in the metal-poor range of [Fe/H]$\le -1.6$.  The conclusion that 
almost all stars with [Fe/H]$\le -1.6$ belong to the halo component is 
supported by no significant change of the $e$ distribution with increasing 
$|z|$ (Figs. 14$a$ and 15).  We have also found no large-scale metallicity 
gradient in the halo in both radial and vertical directions (Figs. 17 and 
18).

Many workers claimed the metal-weak tail of the thick disk component in 
the range of $-1.6<$[Fe/H]$\le -1$ (MFF; Beers \& Sommer-Larsen 1995).  
The fraction of this component $F$ is however found to be smaller than 
previously thought.  The maximum likelihood technique to fit to the observed 
$V_\phi$ distribution provides $F \sim 0.1$ for $-1.6<$[Fe/H]$\le -1$ and 
$F \sim 0.2$ for $-1.4<$[Fe/H]$\le -1$, while $F\sim 0$ for [Fe/H]$\le -1.6$.
We have shown that the distribution of orbital eccentricity provides a 
powerful method for constraining the fraction $F\approx 0.1-0.2$, the mean 
velocity $<V_\phi>_{disk} \approx 195$ km s$^{-1}$, and the vertical extent
$z_{lim}\approx 0.8-1$ kpc of this extra disk component (Figs. $13-15$).
We emphasize that this new approach is effective only if accurate proper 
motions are available by the astrometric satellite like Hipparcos.

We conclude from our results that the extra metal-weak disk which we have 
identified is the metal-weak tail of the rapidly rotating thick disk 
which dominates in the range of [Fe/H]$=-0.6$ to $-1$.  This is therefore 
consistent with the claim by MFF and Beers and Sommer-Larsen (1995), 
although our estimate of $F\approx 0.1-0.2$ is much smaller than theirs.  
Using a full knowledge of the orbital motions of these disk-like stars, 
we have obtained possible indication of the large-scale metallicity gradient 
in the metal-weak tail of the thick disk component (Figs. 19 and 20).

\subsection{Implications for the picture of Galaxy formation}

\subsubsection{The halo component}

Our finding of no significant [Fe/H]$-e$ relation in the range of 
[Fe/H]$\le -1.6$ conflicts with the ELS scenario that the the proto-Galaxy 
underwent the free-fall collapse and the formed stars out of the falling
gas should have eccentric orbits.  The presence of low-$e$ halo stars in
our sample, although comprising only a small fraction, is a key for
understanding how the halo was formed because such low-$e$ stars belong 
to the halo, but not to the rapidly rotating thick disk.  

Our program stars were sampled in the vicinity of the Sun and only about 16\% 
of the sample have eccentricities below $e=0.4$.   As a direct consequence of 
the radially elongated velocity ellipsoid, the $e$ distribution is largely 
skewed towards higher $e$.  On the other hand, the orbital motions of halo
stars sampled at much larger galactocentric distances remain yet undetermined 
because of the lack of accurate measurements of their proper motions.
However, an intriguing result on the velocity distribution in the outer halo 
has been derived by Sommer-Larsen et al. (1994) using the radial velocities 
for their sample of blue horizontal branch stars at $r = 5 - 55$ kpc.  Their 
analyses indicate that the velocity ellipsoid turns out to be tangentially 
anisotropic beyond $r\sim 15$ kpc.  This implies that high angular momentum,
small-$e$ orbits are dominant in the outer halo (bold dashed lines in Figs. 9 
and 10).  Thus, any scenarios for the formation of the Galaxy must explain not
only the $e$ distribution in the solar neighborhood but also the the velocity 
ellipsoid with radial anisotropy transforming into tangential anisotropy with 
increasing galactocentric distance.

If one adopts the currently favored SZ scenario that the halo was assembled 
from merging or accretion of numerous fragments, no correlation between 
kinematical and chemical properties is expected, because each fragment, 
presumably a gas-rich or gas-poor dwarf-type galaxy, has its own chemical 
history.  The SZ scenario is thus successful in explaining no [Fe/H]$-e$
relation and no global metallicity gradient derived from the halo stars 
observed near the Sun.   It is also consistent with a wide age spread in 
globular clusters as well as in field stars, because star formation in each
fragment proceeds independently.  

We proceed to ask whether the SZ scenario is furthermore consistent with the 
$e$ distribution in the solar neighborhood and the change of the velocity 
ellipsoid with increasing galactrocentric distance.  A process of merging or 
accretion of dwarf-type galaxies involves dynamical friction which reduces 
its orbital radius (e.g. Quinn et al. 1993).   At some radius below which the
mean density of the fragment is exceeded by the mean density of the Galaxy, 
the fragment is tidally disrupted and the debris is dispersed to constitute 
the stellar halo.  Since the dynamical friction tends to circularize the orbit
of the fragment, the orbits of remnant stars are weighted in favor of small 
$e$.  This indicates that the velocity ellipsoid becomes more tangentially 
anisotropic at smaller galactocentric distance, which is opposite to the
observed trend.   Although detailed numerical simulations of modeling a number
of accreting events are to be explored, the above simple argument implies that
the SZ scenario seems unlikely to reproduce the kinematical properties of halo 
stars.

An alternative scenario of the formation of the Galaxy has been proposed by 
Sommer-Larsen and Christensen (1989) to explain the change of the velocity 
ellipsoid with galactrocentric distance.  When the proto-Galactic overdense 
region started to collapse out of cosmological expansion, large fluctuations 
developed within the mixture of gas and dark matters made the gas heated up 
to the virial temperature of about $10^6$ K (which is typical of the Galaxy).
This virialized system is largely pressure-supported inside the virial 
radius.  Ensembles of gas clouds are isotropically moving at each radius, 
and dissipative cloud-cloud collisions then induces formation of halo stars.
The collision rate is orbit-dependent.   For example, clouds having more 
radially eccentric orbits encounter with more clouds in denser inner parts 
of the Galaxy, so that such clouds may never return to the radius from 
which the orbital motions start.  Thus, this mechanism favors the survival 
of systematically more circular orbits at larger radii, which agrees with 
the kinematical properties of halo stars.

This scenario was further investigated by Theis (1996) performing numerical 
simulations of collapsing dissipative cloud system.  He has successfully 
obtained the tangentially elongated velocity ellipsoid for survived clouds 
after the dissipative collapse.  It is however yet unexplored whether stars 
formed by this mechanism have the same kinematical and chemical properties 
as observed.  Specifically, since the mechanism involves gaseous dissipation 
over several free-fall time scales, a large-scale metallicity gradient may
appear in the stellar system.  In this respect, the effects of energy feedback
from massive stellar winds and supernovae explosions to the surrounding gas 
may play an important role in suppressing rapid gaseous dissipation and  
smearing out any metallicity gradient by rapid mixing of heavy elements in
the gas.

The more realistic picture is in between the above scenarios.  
The currently favored cold-dark-matter scenario of galaxy formation indicates 
that the initial density fluctuations in the early Universe have larger 
amplitudes on smaller scales (e.g., Padmanabhan 1993).  Hence, the initial 
overdense regions that end up with giant galaxies like our own contain 
larger density fluctuations on sub-galactic scales.  In a collapsing 
protogalaxy these small-scale fluctuations develop to numerous fragments 
which interact together via gravitational force (Katz \& Gunn 1991).   
Due to torque among fragments or direct merging, angular momentum is 
transferred from inner to outer regions of the system.  
Since star formation and chemical evolution differently progress in each 
fragment, one might expect a wide age spread and no metallicity gradient in 
the final stellar system.  This is indeed indistinguishable from the SZ 
scenario. If halo stars are formed via inelastic, anisotropic collisional
processes between fragments, the kinematics of such stars may well accord with
the observed transition of the velocity ellipsoid from the solar neighborhood
to the outer halo (Sommer-Larsen \& Christensen 1989). 

Some of the small density contrasts that have gained systematically higher 
angular momentum in the course of cosmological expansion may have slowly 
fallen to the system after most parts of the system were settled.  These 
delayed accretion may explain the reported indications of relatively young 
stars (Rodgers et al. 1981) and retrograde-orbit stars (Majewski 1992) in 
the outer halo, which have been regarded as a direct evidence of accretion.
It is indeed of great importance to investigate this scenario in more detail, 
by exploring high-resolution simulations of collapsing galaxy combined with 
star formation and chemical evolution, in order to fully understand the 
kinematical and chemical properties of the halo reported in the present work.

\subsubsection{The thick disk component}

How the disk with a large vertical scale height was formed is also enigmatic
(e.g., Majewski 1993).  One leading scenario is that the disk was heated by 
the merging of satellites with the preexisting thin disk (Quinn et al. 1993). 
Satellite orbits were decayed and circularized into the disk plane, and then 
fallen towards the center of the disk.  The disk stars were spread out by 
the merging, and the aftermath was reported to be similar to the observed 
spatial structure and kinematical properties of the thick disk component. 
According to this scenario the thin disk was formed after a major merger
event. 
Therefore, timing of this merger event is severely constrained by the presence
of the presently observed thin disk with a vertical scale height of 350 pc.
An alternative scenario is that the thick disk may have formed in a
dissipative manner after the major parts of the halo formation were 
completed (e.g., Larson 1976; Burkert et al. 1992; Burkert \& Yoshii 1996).
Contraction of the disk either occurred in a pressure-supported manner 
because of the energy feedback or rapidly progressed into the thin disk 
because of the efficient line cooling.

One of the possible observational clues to discriminate these scenarios lies 
in the fraction of the thick disk in the metal-poor range of [Fe/H]$\le -1.6$.
In the merger scenario, since the mechanism relies on both the preexisting old
disk and merging satellites having different chemical histories, the aftermath
of the merger may contain numerous metal-poor stars.  On the contrary, in the 
dissipative-collapse scenario, since the gas that forms the thick disk is 
already enriched by metal ejection of halo stars, only few metal-poor 
stars should be observed in the thick disk.  Our finding of essentially no 
thick-disk stars in the range of [Fe/H]$\le -1.6$ appears to support the 
latter scenario.

Another clue to clarify the formation of the thick disk is to examine whether
a large-scale metallicity gradient exists.  The merger scenario may envisage
no metallicity gradient, whereas the dissipative contraction of the disk may 
involve the smooth spatial variation of metallicity in stars.  No consensus 
has ever met on the observational evidence of metallicity gradient in the 
thick disk (Majewski 1993).  
However, if our finding of non-negligible metallicity gradient in the 
metal-weak disk is the case, it is possible to deduce that the contraction 
of the halo into the thick disk occurred in a dissipative manner just after 
the major parts of the halo formation were completed.

Before concluding definitely, it is required to assemble the data of more 
stars having accurate distances and proper motions.  The method that we have 
developed here based on the eccentricity distribution of orbits may be useful 
for examining whether the thick disk has a significant metal-weak tail as 
well as a global metallicity gradient.   More elaborate modelings are needed 
to further clarify the physical connection between the halo and the thick disk
and to propose what observation will be most definite discriminator of the
scenario of the formation of the Galaxy.

\acknowledgments

We are grateful to H.~Saio, T.~Tsujimoto, M.~Miyamoto for useful discussions.
This work has been supported in part by the Grand-in-Aid for
Scientific Research (08640318, 09640328) and COE Research (07CE2002)
of the Ministry of Education, Science, and Culture in Japan.

\appendix
\section{The ELS model for the gravitational potential}

The ELS potential as a function of galactocentric distance $R$ in the plane
is given by
\begin{equation}
\Phi_{ELS}(R) = - \frac{GM}{b+(R^2+b^2)^{1/2}} \ ,
\end{equation}
where $M$ is the total mass of the disk and $b$ is the scale length. In the
centrifugal equilibrium of the disk, the circular velocity $V_c$ is given by
$V_c(R) = [Rd\Phi_{ELS}/dR]^{1/2}$.

The values of $b$ and $M$ can be evaluated from the Oort constants ($A$,$B$)
and the circular velocity $V_{\odot}$ at $R_{\odot}$ , i.e.,
\begin{eqnarray}
-\frac{A+B}{A-B} &=& \frac{1+2q-q^2}{2q^2}  \\
V_c(R_\odot) &=& V_\odot       \ ,
\end{eqnarray}
where $q \equiv [(R_\odot/b)^2 + 1]^{1/2}$. For $A=15$ km s$^{-1}$ kpc$^{-1}$
and $B=-10$ km s$^{-1}$ kpc$^{-1}$, eq. (A2) gives $q=3.77$ and thus
$b = R_\odot/3.65$ kpc. Equation (A3) then reads $(GM/b)^{1/2}=2.54V_\odot$.
ELS adopted $R_\odot=10$ kpc and $V_\odot=250$ km s$^{-1}$, thereby
$b=2.74$ kpc and $(GM/b)^{1/2}=635$ km s$^{-1}$. If $R_\odot=8$ kpc and
$V_\odot=220$ km s$^{-1}$ as adopted by Carney et al. (1990), we obtain
$b=2.19$ kpc and $(GM/b)^{1/2}=559$ km s$^{-1}$.

In the present work, we use the escape velocity $V_{esc}$ near the Sun as an
alternative constraint. The definition $V_{esc}=[2|\Phi_{ELS}(R_\odot)|]^{1/2}$
then reads
\begin{equation}
1 - \left( \frac{\sqrt{2}V_\odot}{V_{esc}} \right )^2 = \frac{1}{q}  \ ,
\end{equation}
instead of eq. (A2). For $V_\odot=220$ km s$^{-1}$ and $V_{esc}=450$
km s$^{-1}$, we obtain $q=1.92$, thus $b=R_\odot/1.63 =5.2$ kpc. 
Equation (A3) then reads $(GM/b)^{1/2}=2.48V_\odot =545$ km s$^{-1}$
for $V_\odot=220$ km s$^{-1}$. This model is characterized by the larger
$b$ than previous ones to accord with the large $V_{esc}$ as observed
near the Sun.

\clearpage

\clearpage
\begin{center}
TABLE 2  \\
\medskip

L{\sc iterature} S{\sc ources}

\medskip
\begin{tabular}{lc}
\hline\hline 
Source          & Code    \\
\hline
{\it distances/metal abundances}    &  `DA'     \\
\hline
Anthony-Twarog \& Twarog 1994       &  1        \\
                                    &  1s $^a$  \\
Bond 1980                           &  2        \\
                                    &  2s $^a$  \\
Layden 1994                         &  3        \\
Layden 1996                         &  4        \\
\hline
{\it radial velocities}             &  `V'  \\
\hline
Bond 1980                           &  1    \\
Carney \& Latham 1986               &  2    \\
Norris, Bessell \& Pickles 1985     &  3    \\
Barbier-Brossat 1989                &  4    \\
Wilson 1953                         &  5    \\
Evans 1978                          &  6    \\
Griffin et al. 1982                 &  7    \\
Papers quoted by Bond 1980          &  8    \\
Layden 1994                         &  9    \\
\hline
{\it previous results of proper motions}  & `P' \\
\hline
Lick Northern Proper Motion Catalogue    &  1  \\
Hipparcos Input Catalogue                &  2  \\
Wan, Mao \& Ji 1980                      &  3  \\
\hline
\end{tabular}
\begin{flushleft}
\hspace*{2cm}$^a$ Spectroscopic abundances compiled by Anthony-Twarog
\& Twarog (1994).
\end{flushleft}
\end{center}

\clearpage
\begin{center}
TABLE 3  \\
\medskip

M{\sc ean} V{\sc elocities} {\sc and} V{\sc elocity}
D{\sc ispersions} {\sc of} {\sc the} S{\sc ample} S{\sc tars}

\medskip
\begin{tabular}{lrrrrccc}
\hline\hline 
[Fe/H]          & $N$  &
$<U>$ & $<V>$ & $<W>$  &
$\sigma_U$      &  $\sigma_V$  &  $\sigma_W$      \\
(dex)           &      &
(km/s) & (km/s) & (km/s)  &
(km/s)          & (km/s)       & (km/s)           \\
\hline
$+0.1$ to $-0.4$ & 4  &
  14$\pm$7      & $-$21$\pm$8   &  9$\pm$9      &
  40$\pm$16     &    23$\pm$10  & 31$\pm$13    \\
$-0.4$ to $-1.0$ & 13  &
  31$\pm$11    &  $-$78$\pm$15  & $-$11$\pm$17  &
  84$\pm$17    &     86$\pm$18  &  64$\pm$13   \\
$-1.0$ to $-1.6$ & 69  &
 $-$32$\pm$15  & $-$187$\pm$16  &   0$\pm$13    &
  154$\pm$13   &  100$\pm$9     &  94$\pm$8    \\
$\le -1.6$       & 124 &
   16$\pm$18   & $-$217$\pm$21  & $-$10$\pm$12  &
  161$\pm$10   &  115$\pm$7     &  108$\pm$7   \\
$\le -1.8$       & 93  &
    7$\pm$18   & $-$216$\pm$22  & $-$14$\pm$11  &
  160$\pm$12   &  119$\pm$9     &  108$\pm$8   \\
$\le -2.0$       & 64  &
 $-$1$\pm$19   & $-$217$\pm$24  & $-$20$\pm$11  &
  159$\pm$14   &  117$\pm$10    &  111$\pm$10  \\
\hline
\end{tabular}
\end{center}

\bigskip\bigskip\bigskip
\begin{center}
TABLE 4  \\
\medskip

R{\sc otational} P{\sc roperties} {\sc of} {\sc the} S{\sc ample}
S{\sc tars}

\medskip
\begin{tabular}{lcrrcr}
\hline\hline 
[Fe/H]     &  $<$[Fe/H]$>$  & $N$  &
$<V_\phi>$ & $\sigma_\phi$  & $<V_\phi>/\sigma_\phi$  \\
(dex)      &  (dex)         &      &
(km/s)     & (km/s)         &                         \\
\hline
$+0.10$ to $-0.50$ & $-0.20$ &  5 &
 $205\pm 7$ & $ 26\pm 9$ & $ 7.97\pm2.83$ \\
$-0.50$ to $-0.90$ & $-0.70$ &  9 &
 $172\pm13$ & $ 48\pm12$ & $ 3.61\pm0.94$ \\
$-0.90$ to $-1.28$ & $-1.09$ & 22 &
 $ 56\pm11$ & $ 84\pm13$ & $ 0.66\pm0.17$ \\
$-1.28$ to $-1.45$ & $-1.37$ & 21 &
 $ 39\pm16$ & $108\pm17$ & $ 0.36\pm0.16$ \\
$-1.45$ to $-1.56$ & $-1.50$ & 18 &
 $ 18\pm17$ & $107\pm18$ & $ 0.17\pm0.17$ \\
$-1.56$ to $-1.75$ & $-1.65$ & 36 &
 $ -7\pm19$ & $ 97\pm12$ & $-0.08\pm0.20$ \\
$-1.75$ to $-1.95$ & $-1.85$ & 26 &
 $ 27\pm17$ & $119\pm17$ & $ 0.22\pm0.14$ \\
$-1.95$ to $-2.30$ & $-2.12$ & 33 &
 $ 12\pm20$ & $115\pm14$ & $ 0.10\pm0.17$ \\
$-2.30$ to $-3.01$ & $-2.65$ & 38 &
 $-11\pm19$ & $122\pm14$ & $-0.09\pm0.15$ \\
\hline
\end{tabular}
\end{center}

\clearpage
\begin{center}
TABLE 5  \\
\medskip

P{\sc arameters} {\sc of} {\sc the} M{\sc etal} W{\sc eak}
T{\sc hick} D{\sc isk} {\sc at} $|z|<1$ {\sc kpc}

\medskip
\begin{tabular}{cccccc}
\hline\hline 
[Fe/H]          &  Stars  & $N$  &
$<V_\phi>_{disk}$ & $\sigma_{\phi,{disk}}$  &  $F$       \\
(dex)           &         &      &
(km/s)            & (km/s)                 &            \\
\hline
  \multicolumn{6}{c}{$<V_\phi>_{disk}$ (varied)}      \\
\hline
$-1.0$ to $-1.6$& Red giants & 23 &  120 &  50  &  0.34  \\
                & RR Lyraes  & 46 &  117 &  72  &  0.33  \\
                & Both stars & 69 &  113 &  64  &  0.33  \\
$-1.0$ to $-1.5$& Red giants & 14 &  129 &  51  &  0.36  \\
                & RR Lyraes  & 33 &  120 &  73  &  0.34  \\
                & Both stars & 47 &  118 &  66  &  0.34  \\
$-1.0$ to $-1.4$& Red giants & 11 &  137 &  56  &  0.37  \\
                & RR Lyraes  & 24 &  145 &  64  &  0.39  \\
                & Both stars & 35 &  140 &  62  &  0.38  \\
\hline
  \multicolumn{6}{c}{$<V_\phi>_{disk}$ (fixed at 195 km s$^{-1}$)}  \\
\hline
$-1.0$ to $-1.6$& Red giants & 23 &  195 &  44  &  0.09  \\
                & RR Lyraes  & 46 &  195 &  33  &  0.12  \\
                & Both stars & 69 &  195 &  36  &  0.09  \\
$-1.0$ to $-1.5$& Red giants & 14 &  195 &  41  &  0.18  \\
                & RR Lyraes  & 33 &  195 &  50  &  0.15  \\
                & Both stars & 47 &  195 &  41  &  0.12  \\
$-1.0$ to $-1.4$& Red giants & 11 &  195 &  34  &  0.26  \\
                & RR Lyraes  & 24 &  195 &  54  &  0.28  \\
                & Both stars & 35 &  195 &  41  &  0.23  \\
\hline
\end{tabular}
\end{center}

\clearpage
\begin{center}
TABLE 6  \\
\medskip

F{\sc raction} {\sc of} S{\sc mall} P{\sc lanar}-$e$ O{\sc rbits}
{\sc for} {\sc the} ELS P{\sc otential} $^a$

\medskip
\begin{tabular}{lcc}
\hline\hline 
Case               &   \multicolumn{2}{c}{fraction with $e<0.4$}   \\
\cline{2-3}
                   &  observation  & prediction $^b$   \\
                   &    (\%)       &    (\%)           \\
\hline
\multicolumn{3}{c}{$R_{\odot}=8.5$ kpc, $V_{\odot}=220$ km/s}  \\
\hline
$V_{esc}=450$ km/s  &    12.7    &   13.9        \\
$V_{esc}=400$ km/s  &    10.5    &   11.4        \\
$V_{esc}=500$ km/s  &    17.9    &   15.8        \\
\hline
\multicolumn{3}{c}{no constriant on $V_{esc}$}   \\
\hline
$R_{\odot}=8$ kpc, $V_{\odot}=220$ km/s $^c$  &  5.3  &  8.7  \\
$R_{\odot}=10$ kpc, $V_{\odot}=250$ km/s $^d$ &  4.5  &  7.6  \\
\hline
\end{tabular}
\begin{flushleft}
\hspace*{3.2cm}$^a$ For [Fe/H]$\le -1.6$. \\
\hspace*{3.2cm}$^b$ For $(\sigma_U, \sigma_V)=(161, 115)$ km s$^{-1}$
 (see the text in detail).\\
\hspace*{3.2cm}$^c$ Parameters adopted by Carney et al (1990).\\
\hspace*{3.2cm}$^d$ Parameters adopted by ELS and NBP. \\
\end{flushleft}
\end{center}

\clearpage
\begin{center}
TABLE 7  \\
\medskip

M{\sc etallicity} {\sc vs.} A{\sc pogalactic} C{\sc ylindrical}
D{\sc istance}

\medskip
\begin{tabular}{ccccccccc}
\hline\hline 
 & \multicolumn{2}{c}{$V_\phi\le\infty$}  &
   \multicolumn{2}{c}{$V_\phi\le170$km/s} &
   \multicolumn{2}{c}{$V_\phi\le120$km/s} &
   \multicolumn{2}{c}{$V_\phi\le70$km/s}            \\
\cline{2-9}
range in $R_{max}$ & $N$ & $<$[Fe/H]$>$ & $N$ & $<$[Fe/H]$>$    
                   & $N$ & $<$[Fe/H]$>$ & $N$ & $<$[Fe/H]$>$  \\
    (kpc)          &     &    (dex)     &     &   (dex)         
                   &     &    (dex)     &     &   (dex)       \\
\hline
  \multicolumn{9}{c}{Halo candidates}    \\
\hline
 7.0$-$9.0 & 50 &$-$1.59$\pm$0.18& 46 &$-$1.67$\pm$0.17  &
             40 &$-$1.73$\pm$0.17& 31 &$-$1.82$\pm$0.15    \\
 9.0$-$12.0& 79 & $-$1.76   0.15 & 73 & $-$1.84   0.15   &
             59 & $-$1.90   0.15 & 44 & $-$1.95   0.15     \\
12.0$-$18.0& 37 & $-$1.82   0.14 & 31 & $-$1.85   0.14   &
             30 & $-$1.83   0.15 & 24 & $-$1.84   0.14     \\
18.0$-$25.0& 27 & $-$1.81   0.15 & 27 & $-$1.81   0.15   &
             23 & $-$1.80   0.16 & 20 & $-$1.76   0.16     \\
25.0$-$40.0& 14 & $-$1.83   0.18 & 13 & $-$1.80   0.18   &
             12 & $-$1.70   0.19 & 11 & $-$1.69   0.19     \\
\hline
  \multicolumn{9}{c}{MWTD candidates $^a$}    \\
\hline
 7.0$-$9.0 & 22 &$-$1.41$\pm$0.20& 19 &$-$1.56$\pm$0.19  &
             14 &$-$1.62$\pm$0.20&  7 &$-$1.96$\pm$0.14    \\
 9.0$-$12.0& 30 & $-$1.47   0.14 & 24 & $-$1.64   0.14   &
             10 & $-$1.74   0.14 &  4 & $-$2.13   0.13     \\
12.0$-$18.0& 12 & $-$1.89   0.15 &  7 & $-$1.90   0.14   &
              6 & $-$1.84   0.15 &  3 & $-$1.87   0.10     \\
18.0$-$25.0&  5 & $-$1.97   0.11 &  5 & $-$1.97   0.11   &
              1 & $-$2.37   0.05 &  1 & $-$2.37   0.05     \\
25.0$-$40.0&  1 & $-$1.72   0.28 &  1 & $-$1.72   0.28   &
              1 & $-$1.72   0.28 &  1 & $-$1.72   0.28     \\
\hline
\end{tabular}
\begin{flushleft}
\hspace*{1cm}$^a$ With extra constraints of $e\le0.6$ and $|z|<1$ kpc.
\end{flushleft}
\end{center}

\clearpage
\begin{center}
TABLE 8  \\
\medskip

M{\sc etallicity} {\sc vs.} M{\sc aximum} V{\sc ertical} D{\sc istance}

\medskip
\begin{tabular}{ccccccccc}
\hline\hline 
 & \multicolumn{2}{c}{$V_\phi\le\infty$}  &
   \multicolumn{2}{c}{$V_\phi\le170$km/s} &
   \multicolumn{2}{c}{$V_\phi\le120$km/s} &
   \multicolumn{2}{c}{$V_\phi\le70$km/s}            \\
\cline{2-9}
range in $|z_{max}|$ & $N$ & $<$[Fe/H]$>$ & $N$ & $<$[Fe/H]$>$    
                     & $N$ & $<$[Fe/H]$>$ & $N$ & $<$[Fe/H]$>$  \\
    (kpc)            &     &    (dex)     &     &   (dex)         
                     &     &    (dex)     &     &   (dex)       \\
\hline
  \multicolumn{9}{c}{Halo candidates}    \\
\hline
  0.0$-$1.0  & 48 &$-$1.56$\pm$0.15 & 41 &$-$1.74$\pm$0.14  
             & 33 &$-$1.81$\pm$0.13 & 26 &$-$1.84$\pm$0.14  \\
  1.0$-$2.0  & 48 & $-$1.72    0.17 & 44 & $-$1.74   0.17   
             & 38 & $-$1.82    0.17 & 27 & $-$1.92    0.14  \\
  2.0$-$4.0  & 48 & $-$1.82    0.15 & 45 & $-$1.85   0.15   
             & 39 & $-$1.82    0.16 & 32 & $-$1.86    0.16  \\
  4.0$-$8.0  & 36 & $-$1.80    0.16 & 35 & $-$1.80   0.16   
             & 34 & $-$1.78    0.16 & 29 & $-$1.77    0.16  \\
  8.0$-$15.0 & 17 & $-$1.87    0.16 & 16 & $-$1.89   0.16   
             & 15 & $-$1.87    0.17 & 13 & $-$1.83    0.17  \\
 15.0$-$40.0 & 12 & $-$1.91    0.15 & 10 & $-$1.90   0.14   
             &  6 & $-$1.91    0.16 &  4 & $-$1.99    0.15  \\
\hline
  \multicolumn{9}{c}{MWTD candidates $^a$}    \\
\hline
  0.0$-$1.0  & 20 &$-$1.17$\pm$0.17 & 13 &$-$1.54$\pm$0.15  
             &  5 &$-$1.71$\pm$0.14 &  3 &$-$1.73$\pm$0.17  \\
  1.0$-$2.0  & 20 & $-$1.53   0.19  & 17 & $-$1.51   0.20   
             & 12 & $-$1.61   0.21  &  5 & $-$2.18   0.14   \\
  2.0$-$4.0  & 13 & $-$1.80   0.14  & 11 & $-$1.81   0.13   
             &  6 & $-$1.83   0.14  &  3 & $-$2.07   0.08   \\
  4.0$-$8.0  &  7 & $-$1.99   0.14  &  6 & $-$1.99   0.13   
             &  5 & $-$1.93   0.13  &  3 & $-$2.06   0.10   \\
  8.0$-$15.0 &  5 & $-$1.75   0.17  &  4 & $-$1.80   0.18   
             &  3 & $-$1.66   0.21  &  2 & $-$1.71   0.22   \\
 15.0$-$40.0 &  5 & $-$1.85   0.14  &  5 & $-$1.85   0.14   
             &  1 & $-$1.76   0.20  &  0 & ........         \\
\hline
\end{tabular}
\begin{flushleft}
\hspace*{1cm}$^a$ With extra constraints of $e\le0.6$ and $|z|<1$ kpc.
\end{flushleft}
\end{center}

\clearpage
\begin{figure}
\plotone{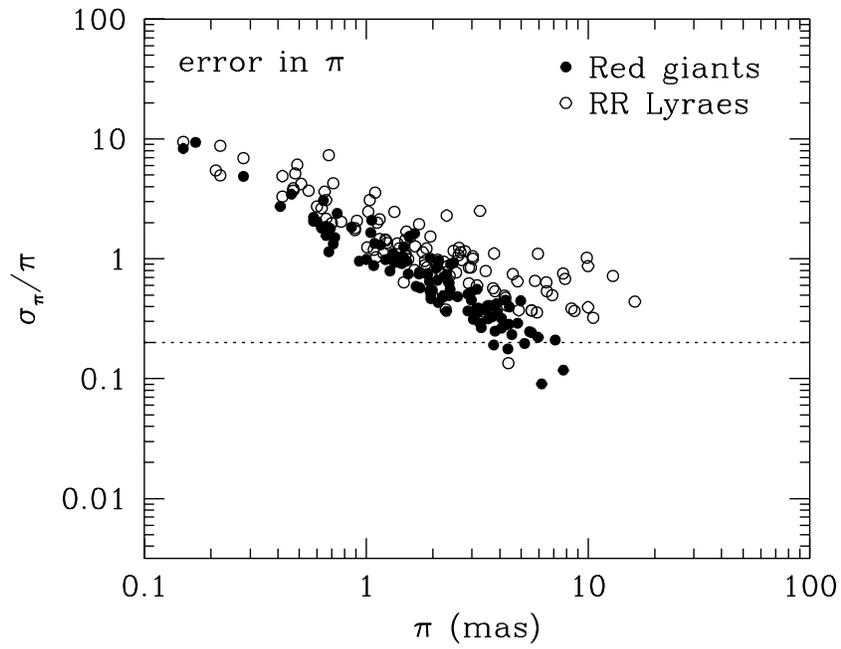}
\caption{
The distribution of relative parallax errors $\sigma_\pi/\pi$. Filled and
open circles denote red giants and RR Lyraes, respectively.
}
\end{figure}

\clearpage
\begin{figure}
\plotone{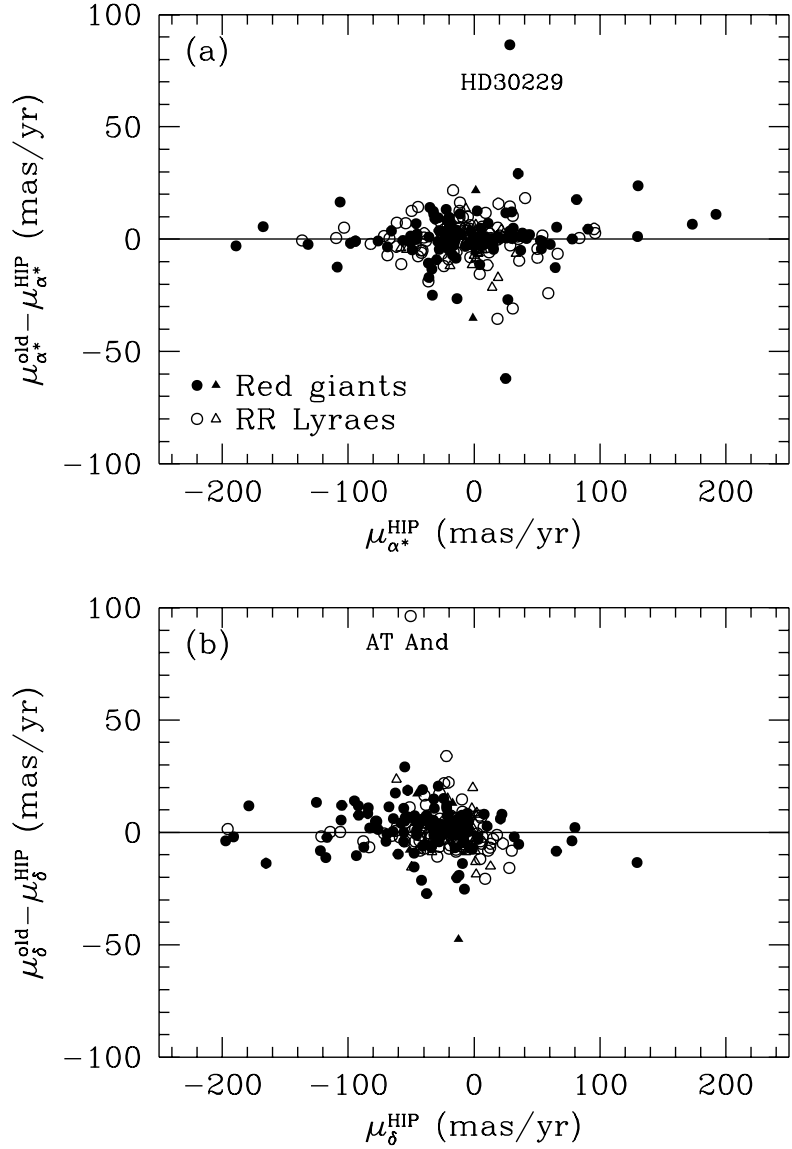}
\caption{
The difference between the previous (superscript 'old') and the Hipparcos
('HIP') measurements for proper motions $\mu_{\alpha^\ast}$ (a) and
$\mu_{\delta}$ (b). Filled and open circles denote red giants and
RR Lyraes, respectively, with small relative errors in proper motions
($|\mu_{\alpha^\ast}| > \sigma_{\mu_{\alpha^\ast}}$ and $|\mu_{\delta}| >
\sigma_{\mu_{\delta}}$), while filled and open triangles are for large errors
($|\mu_{\alpha^\ast}| \le \sigma_{\mu_{\alpha^\ast}}$ or $|\mu_{\delta}| \le
\sigma_{\mu_{\delta}}$).
}
\end{figure}

\clearpage
\begin{figure}
\plotone{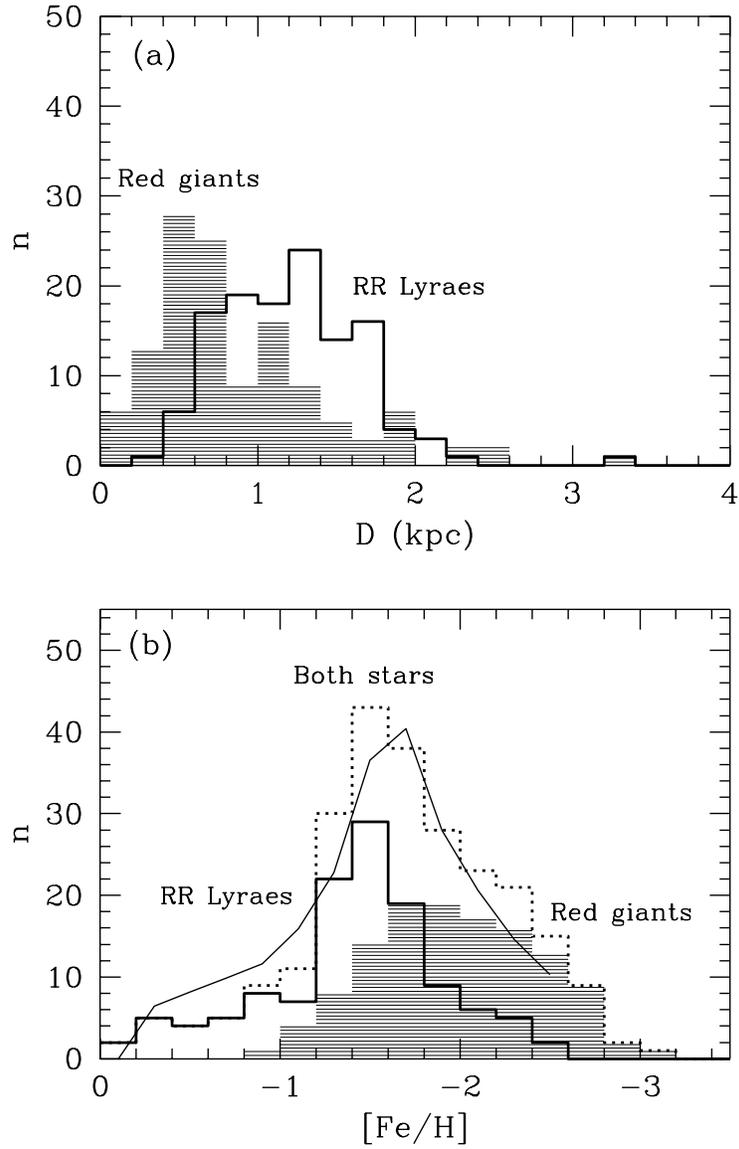}
\caption{
The distributions of distances (a) and metallicities (b) for red giants
(shaded histograms) and RR Lyraes (solid histograms). In panel $b$, dotted
histogram is for both stars, while solid line shows the likely true metallicity
distribution of halo stars derived by Laird et al. (1988).
}
\end{figure}

\clearpage
\begin{figure}
\plotone{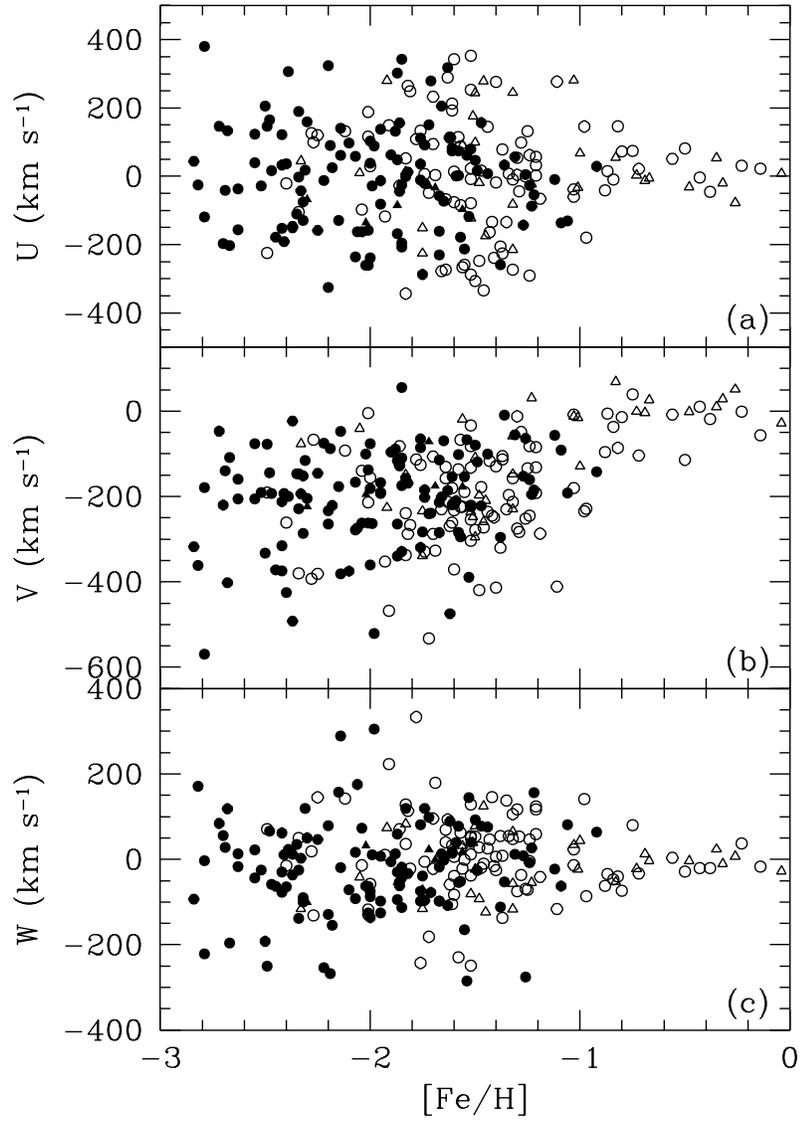}
\caption{
$(U,V,W)$ velocity components versus [Fe/H] for the sample. The symbol
designation is the same as in Fig. 2.
}
\end{figure}

\clearpage
\begin{figure}
\plotone{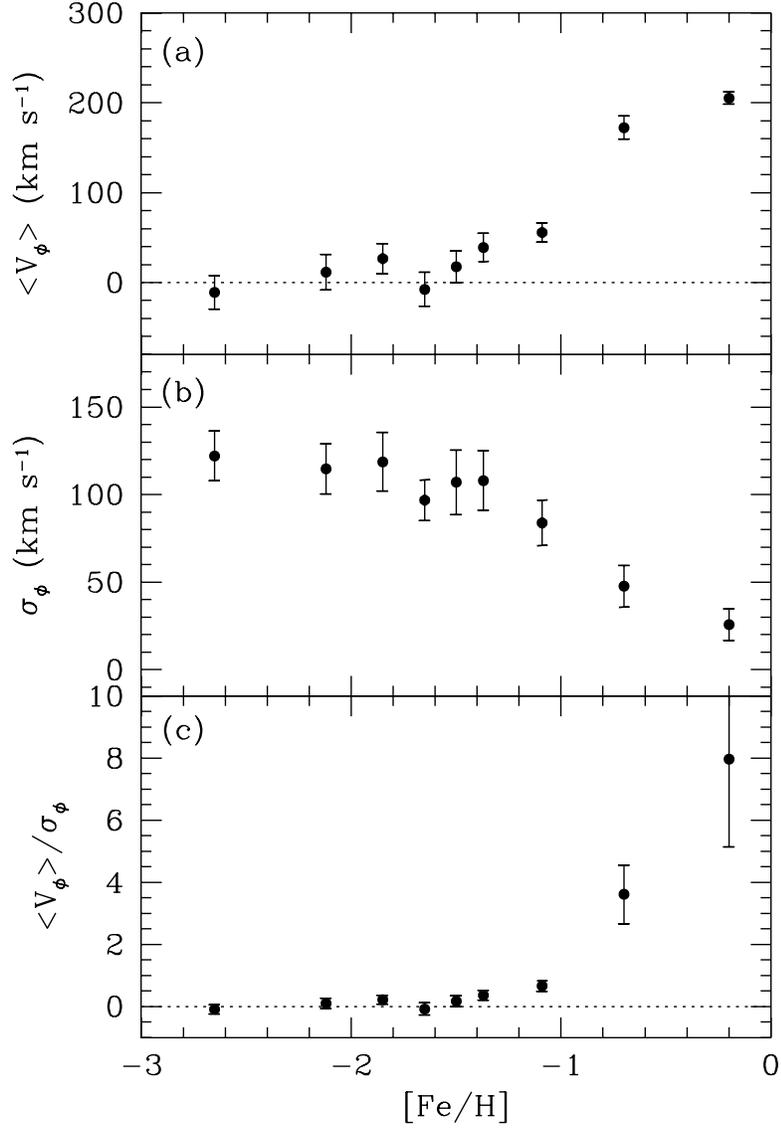}
\caption{
Rotational properties versus [Fe/H] for the sample.
(a) The mean rotation $<V_\phi>$.
(b) The velocity dispersion in $\phi$ direction, $\sigma_\phi$.
(c) The ratio $<V_\phi>/\sigma_\phi$.
}
\end{figure}

\clearpage
\begin{figure}
\plotone{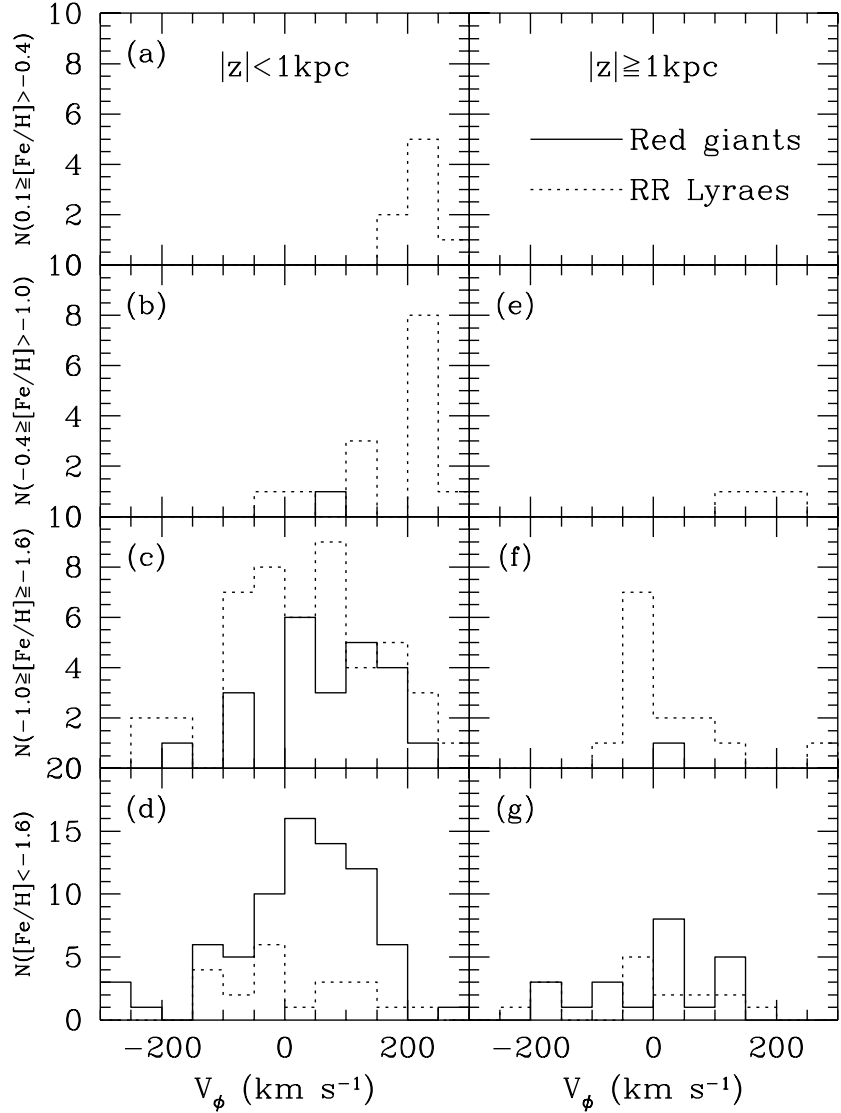}
\caption{
The distributions of $V_\phi$ in various metallicity ranges for red giants
(solid lines) and RR Lyraes (dotted lines). The left and right panels are
for $|z|<1$ kpc and $|z|\ge 1$ kpc, respectively, and the metallicity ranges
are indicated in the labels for vertical axes. Note that in the range
of $-1.6\le$[Fe/H]$\le-1$ (panel $c$), the deviation from a single Gaussian
is less significant than previously reported.
}
\end{figure}

\clearpage
\begin{figure}
\plotone{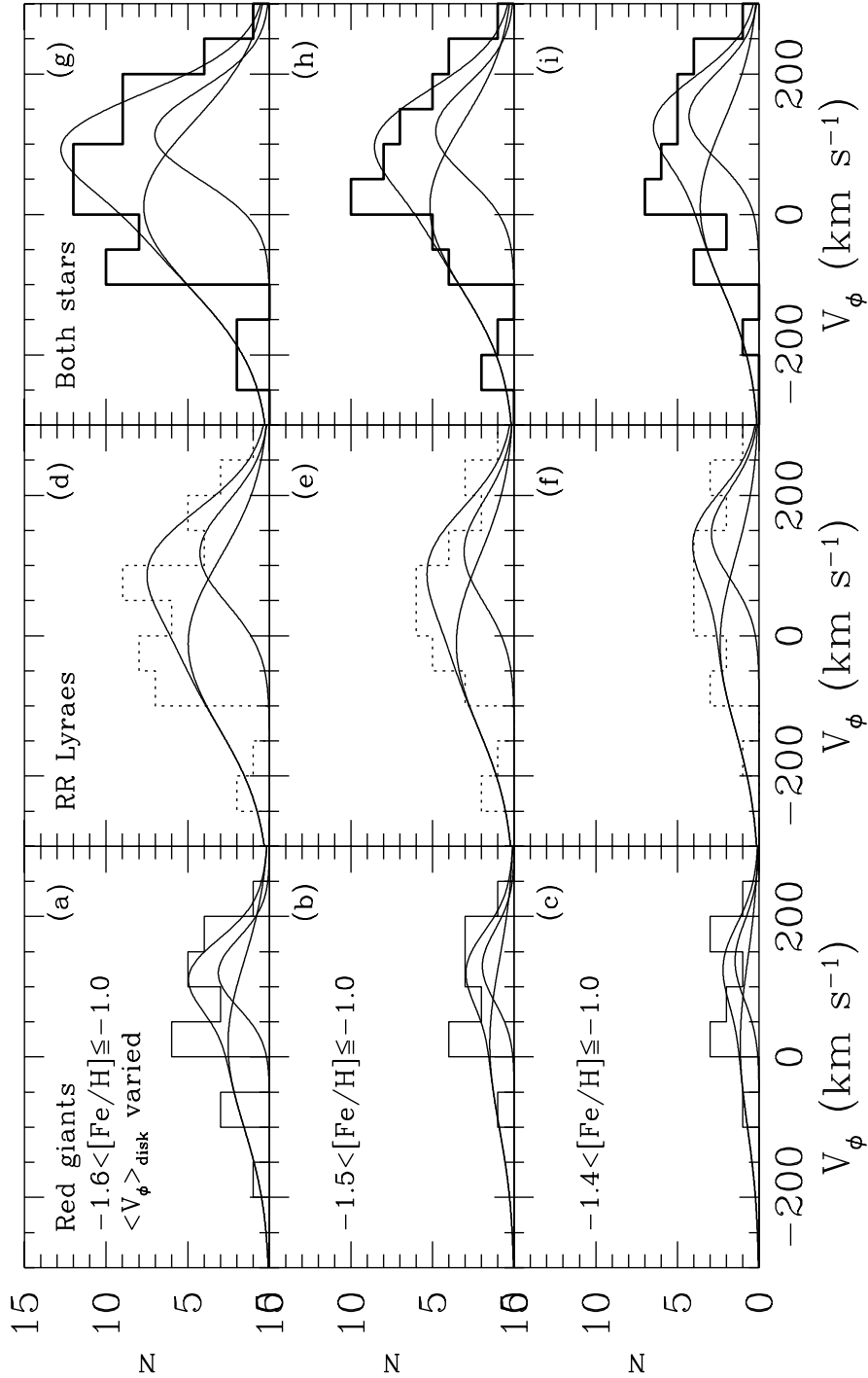}
\caption{
The results of the maximum likelihood method for reproducing the $V_\phi$
distribution at $|z|<1$ kpc, based on a mixture of two Gaussian components
(halo$+$disk). The mean rotation of the disk $<V_\phi>_{disk}$ is one of
the variable parameters for fitting. The metallicity ranges are
$-1.6<$[Fe/H]$\le -1$ (panels $a$, $d$, and $g$),
$-1.5<$[Fe/H]$\le -1$ (panels $b$, $e$, and $h$), and
$-1.4<$[Fe/H]$\le -1$ (panels $c$, $f$, and $i$).
}
\end{figure}

\clearpage
\begin{figure}
\plotone{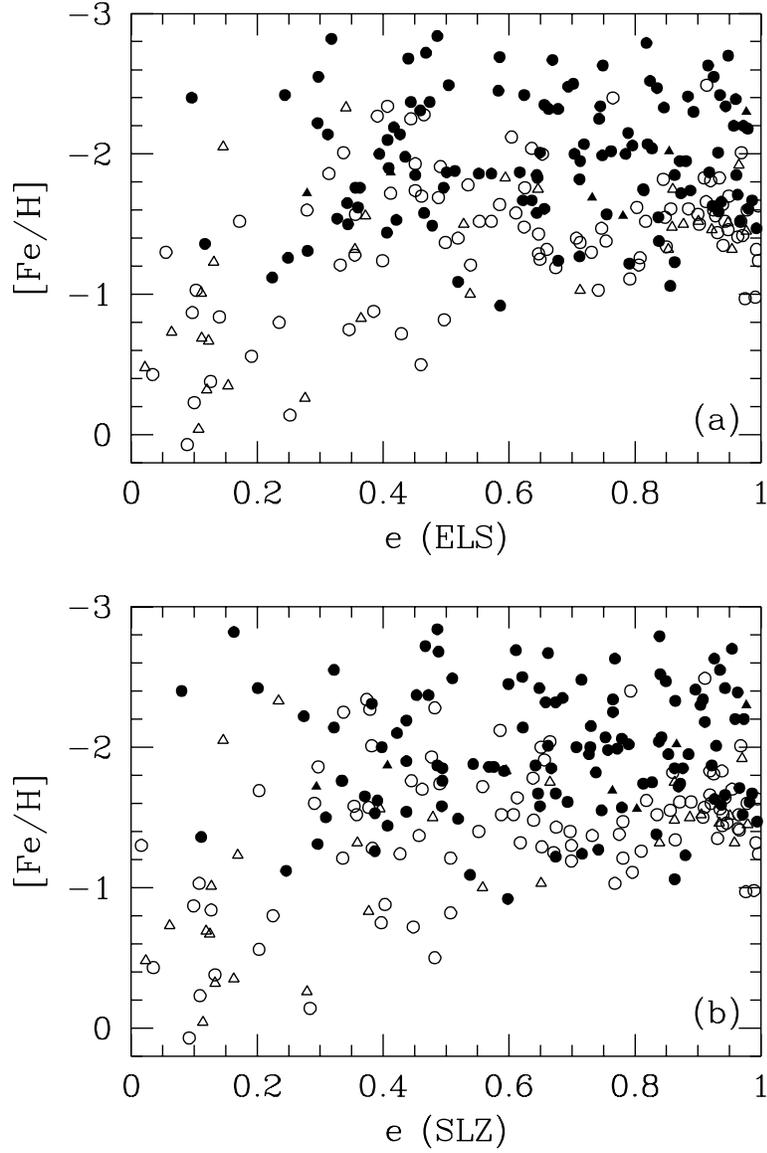}
\caption{
The relation between [Fe/H] and $e$ for the ELS (a) and the SLZ (b)
gravitational potentials. The symbol designation is the same as in Fig. 2.
}
\end{figure}

\clearpage
\begin{figure}
\plotone{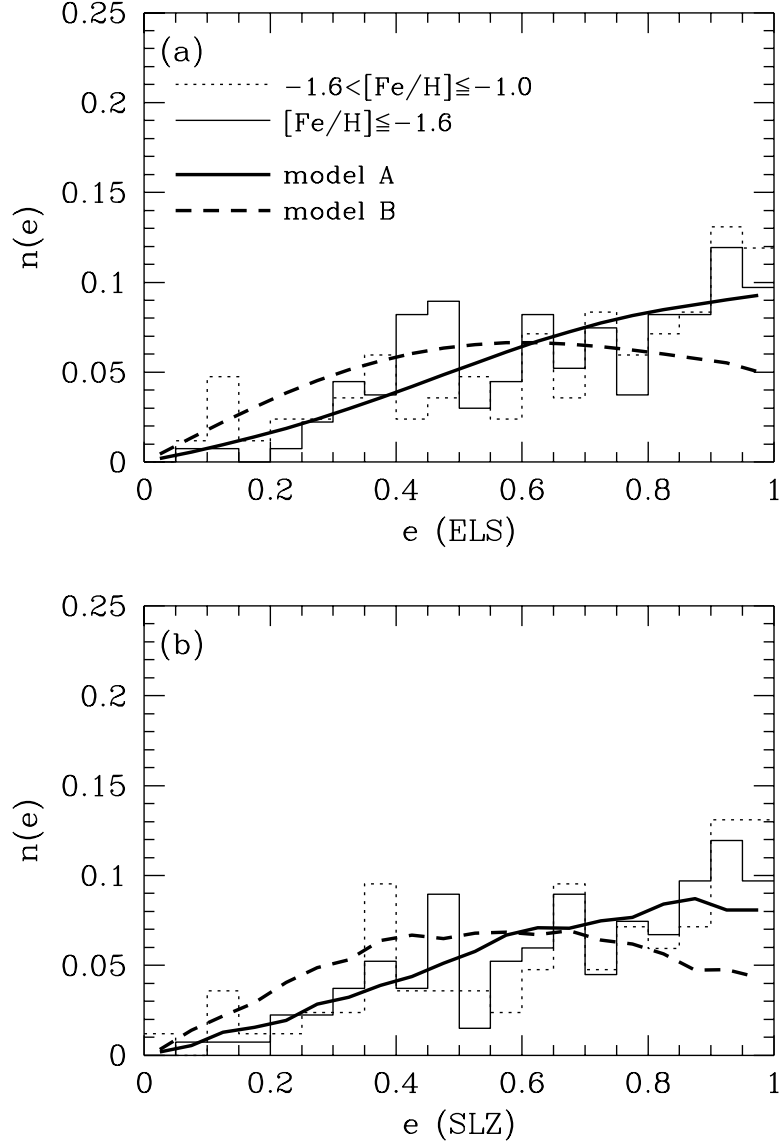}
\caption{
The normalized differential $e$ distribution $n(e)$ for the sample stars with
[Fe/H]$\le -1.6$ (solid histograms) and $-1.6<$[Fe/H]$\le -1$ (dotted
histograms). Bold solid (model A) and bold dashed lines (model B) denote
the model predictions based on a single Gaussian velocity distribution
with a radially anisotropic $(\sigma_U,\sigma_V,\sigma_W)=(161, 115, 108)$
km s$^{-1}$ and a tangentially anisotropic $(115, 161, 108)$ km s$^{-1}$
velocity ellipsoid, respectively. The former velocity ellipsoid is derived
from stars with [Fe/H]$\le -1.6$, whereas the latter one is just for
the purpose of comparison.
}
\end{figure}

\clearpage
\begin{figure}
\plotone{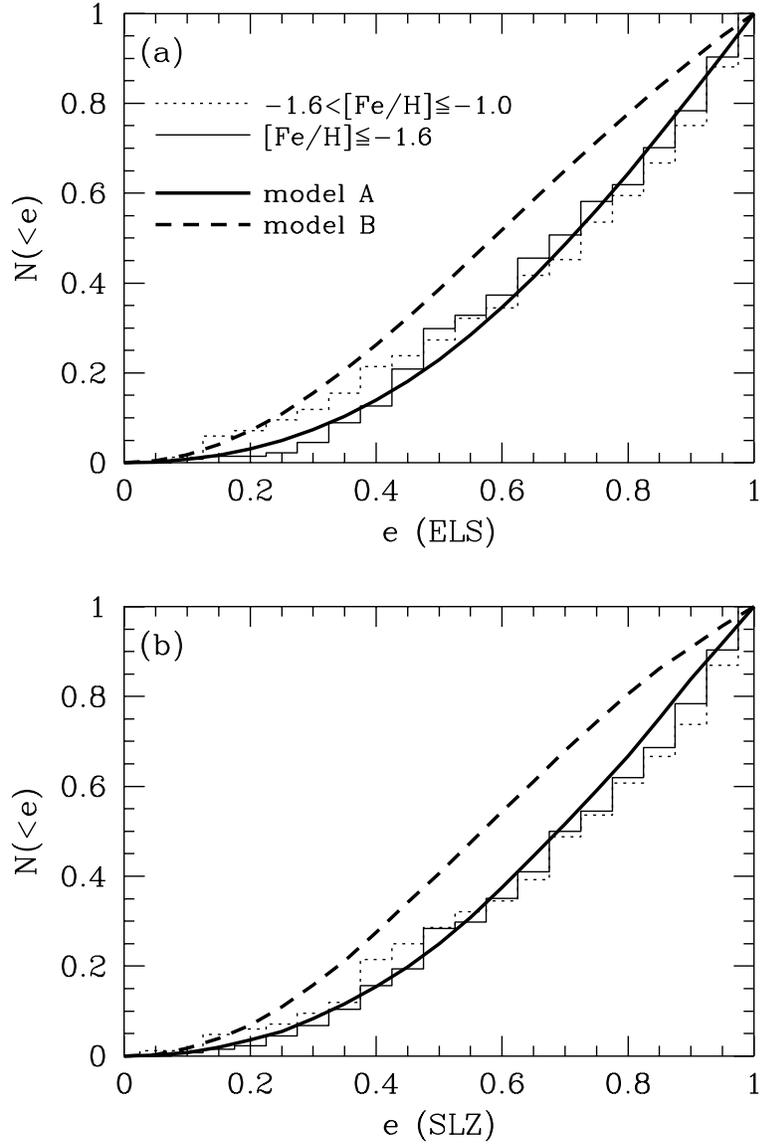}
\caption{
The cumulative $e$ distribution $N(<e)$ for the sample.
Others are the same as for Fig. 9.
}
\end{figure}

\clearpage
\begin{figure}
\plotone{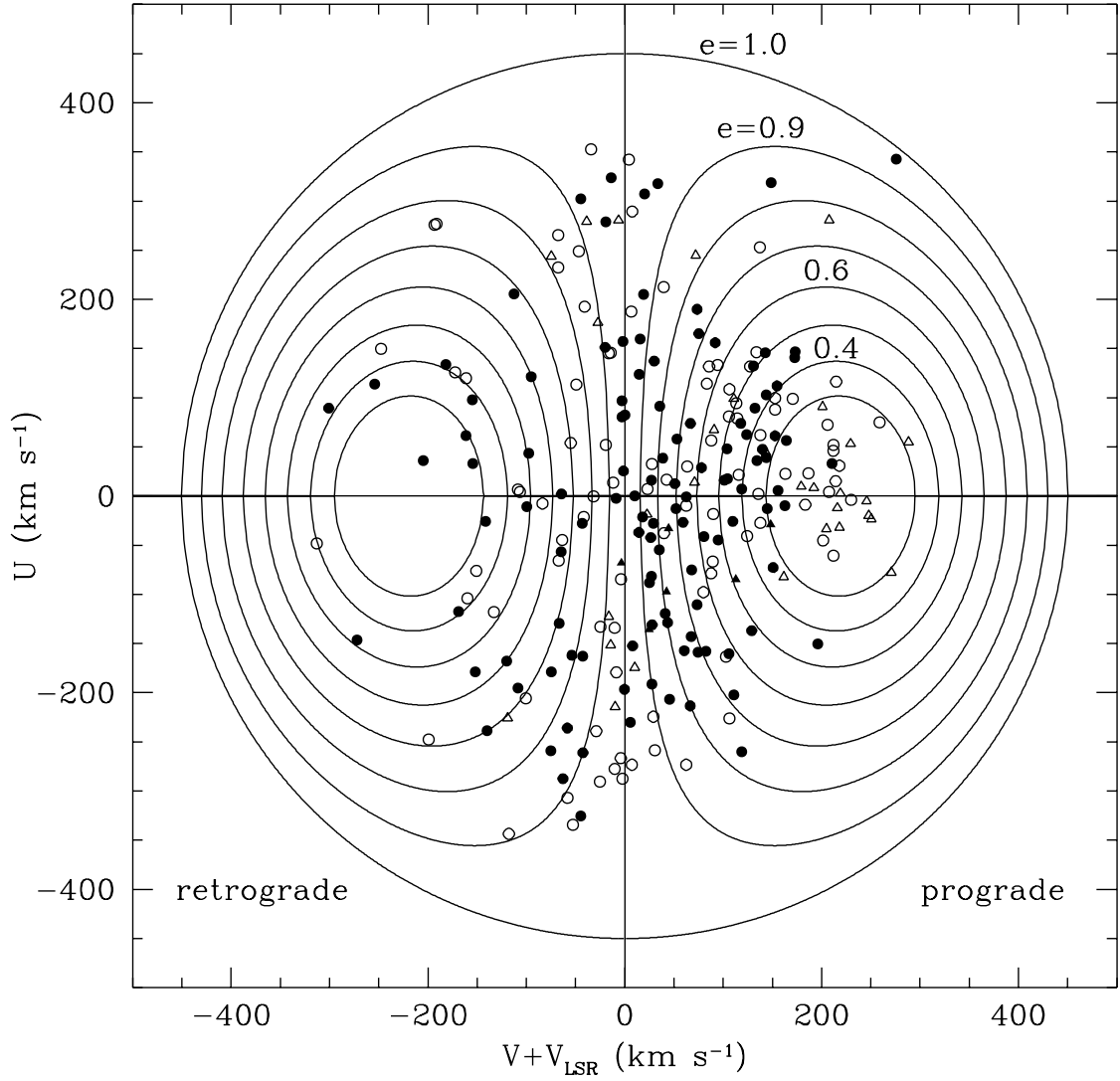}
\caption{
The Bottlinger diagram for the sample stars. The symbol designation is the
same as in Fig. 2. Each curve denotes a locus of constant $e$
derived from the ELS gravitational potential.
}
\end{figure}

\clearpage
\begin{figure}
\plotone{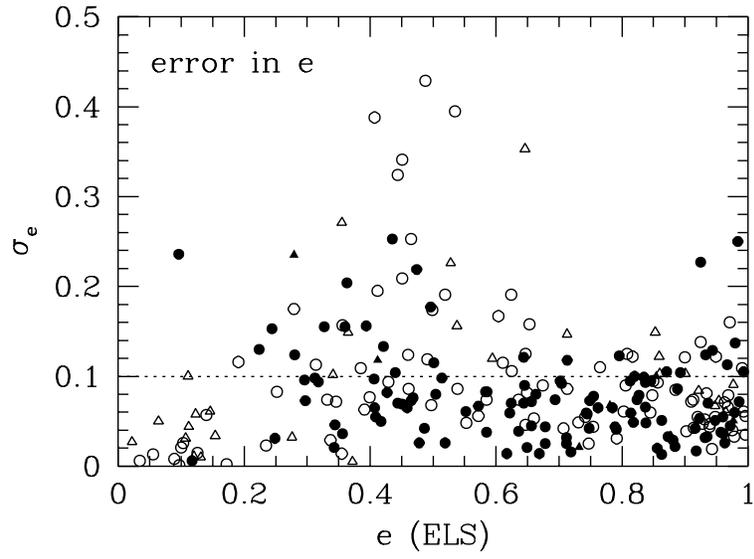}
\caption{
The distribution of errors $\sigma_e$ in $e$ for the ELS gravitational
potential.
}
\end{figure}

\clearpage
\begin{figure}
\plotone{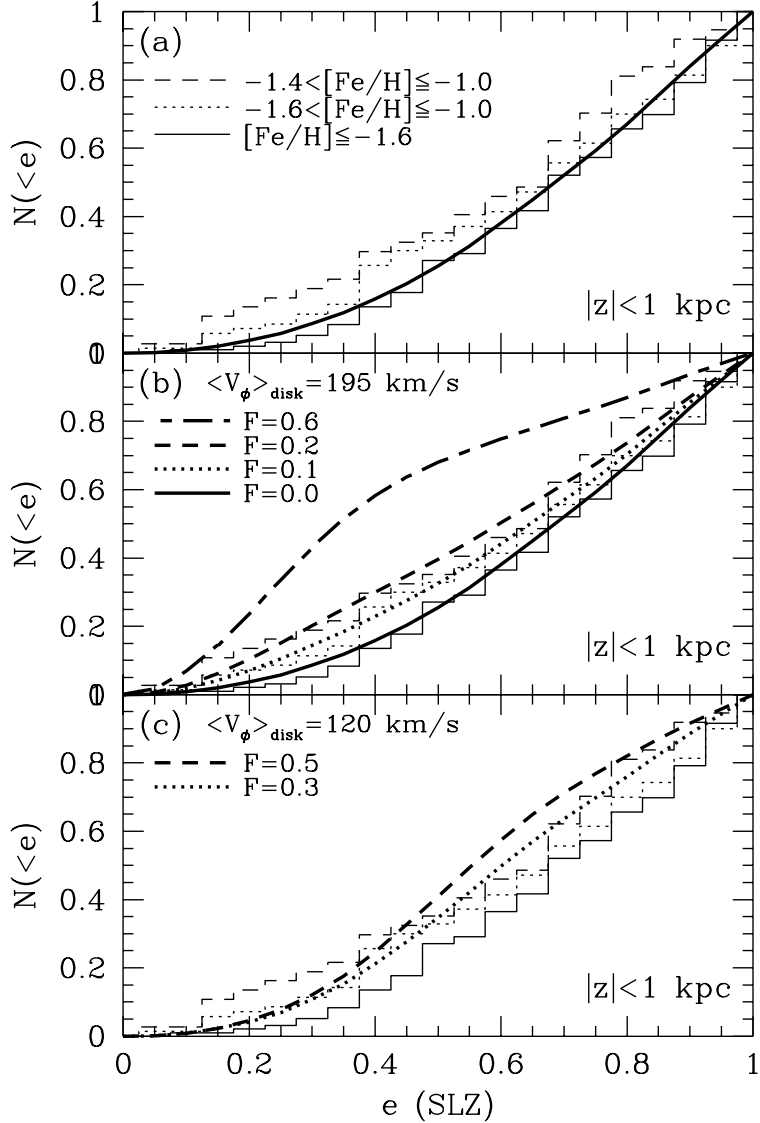}
\caption{
The cumulative $e$ distribution for the stars at $|z|<$1 kpc,
in the metallicity ranges of [Fe/H]$\le -1.6$ (solid histograms),
$-1.6<$[Fe/H]$\le -1$ (dotted histograms), and $-1.4<$[Fe/H]$\le -1$
(dashed histograms). The SLZ gravitational potential is used.
Bold solid line in panel $a$ corresponds to the
model prediction using the velocity ellipsoid
$(\sigma_U,\sigma_V,\sigma_W)=(165, 120, 107)$ km s$^{-1}$
obtained for stars at $|z|<1$ kpc with [Fe/H]$<-1.6$.
Various bold lines in panels $b$ and $c$ denote the model results based on
a mixture of two Gaussian components (thick disk$+$halo). The quantity $F$
denotes the fraction of the thick-disk component. Panel $b$ is for
the mean disk rotation of $<V_\phi>_{disk}=195$ km s$^{-1}$,
while panel $c$ is for $<V_\phi>_{disk}=120$ km s$^{-1}$. 
}
\end{figure}

\clearpage
\begin{figure}
\plotone{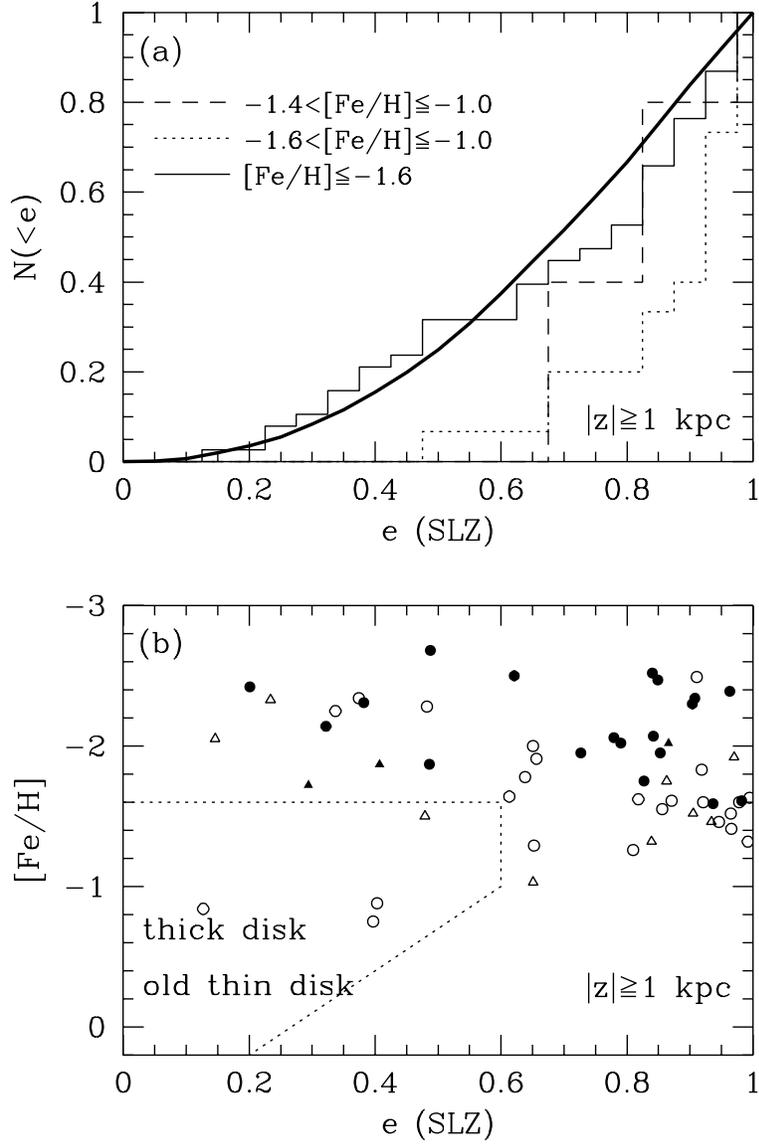}
\caption{
(a) The normalized cumulative $e$ distribution for stars at $|z|\ge$1 kpc.
Bold solid line is the same as that (model A) in Fig. 10. Others are the
same as for Fig. 10.
(b) The relation between [Fe/H] and $e$ at $|z|\ge 1$ kpc for the SLZ
gravitational potential. Note that when comparing with Fig. 8
for all $z$, stars enclosed by dotted lines are selectively excluded by
the constraint $|z|\ge 1$ kpc.
}
\end{figure}

\clearpage
\begin{figure}
\plotone{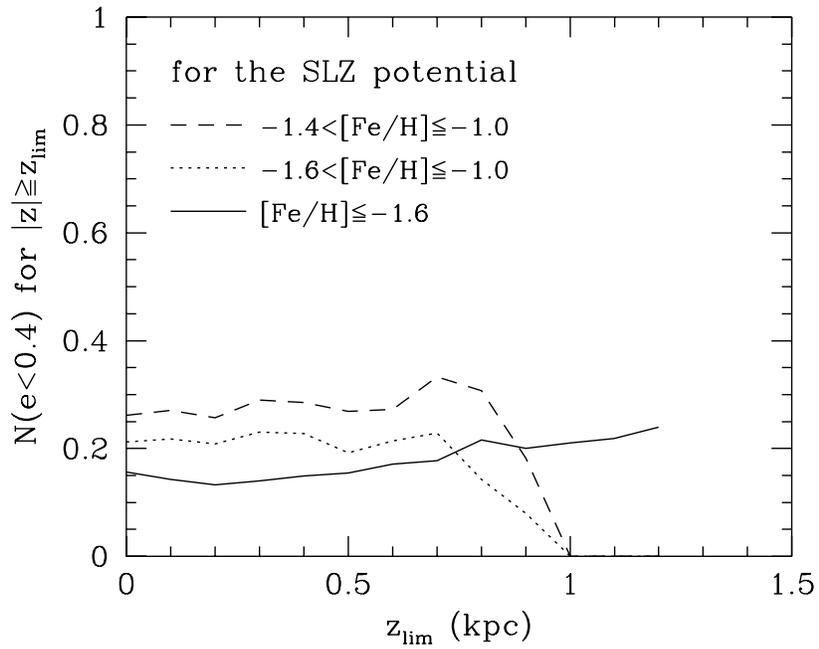}
\caption{
The fraction of stars having $e<0.4$ for $|z|\ge z_{lim}$, as a function of
$z_{lim}$. Solid, dotted, and dashed lines are for [Fe/H]$\le -1.6$,
$-1.6<$[Fe/H]$\le -1$, and $-1.4<$[Fe/H]$\le -1$, respectively.
Note the sharp decrease of the curves at $z_{lim}=0.8-1$ kpc for
the intermediate metallicity range, whereas the curve for [Fe/H]$\le -1.6$
remains essentially unchanged at large $z_{lim}$.
}
\end{figure}

\clearpage
\begin{figure}
\plotone{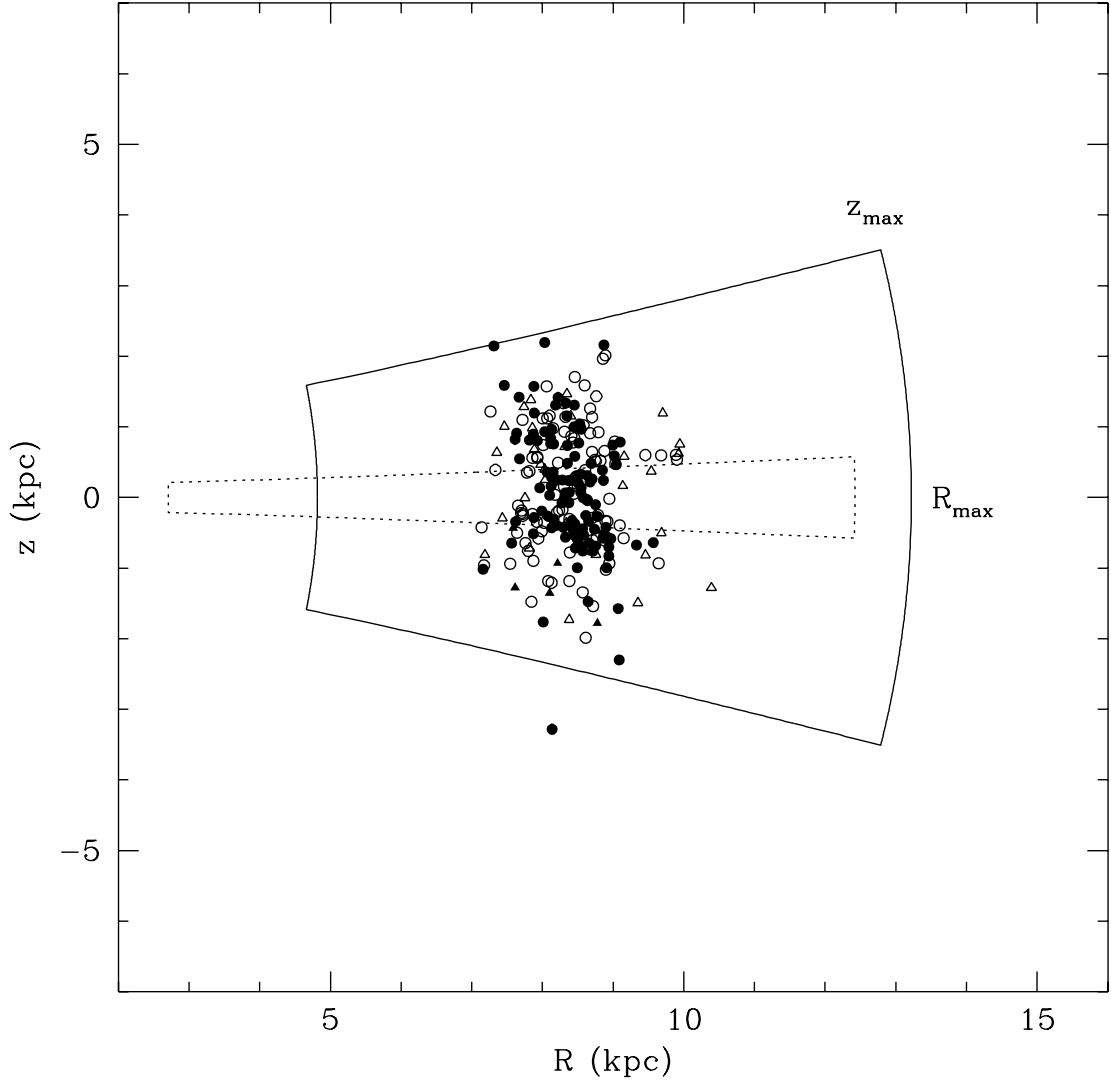}
\caption{
The spatial distribution of the sample stars in $(R,z)$. The area enclosed
by solid lines corresponds to the domain of orbital motions for HIC\#3554
at $(R,z)=(8.58,-0.54)$ kpc, whereas dotted lines are for HIC\#2413
at $(R,z)=(8.62,-0.02)$ kpc. The SLZ gravitational model is used.
}
\end{figure}

\clearpage
\begin{figure}
\plotone{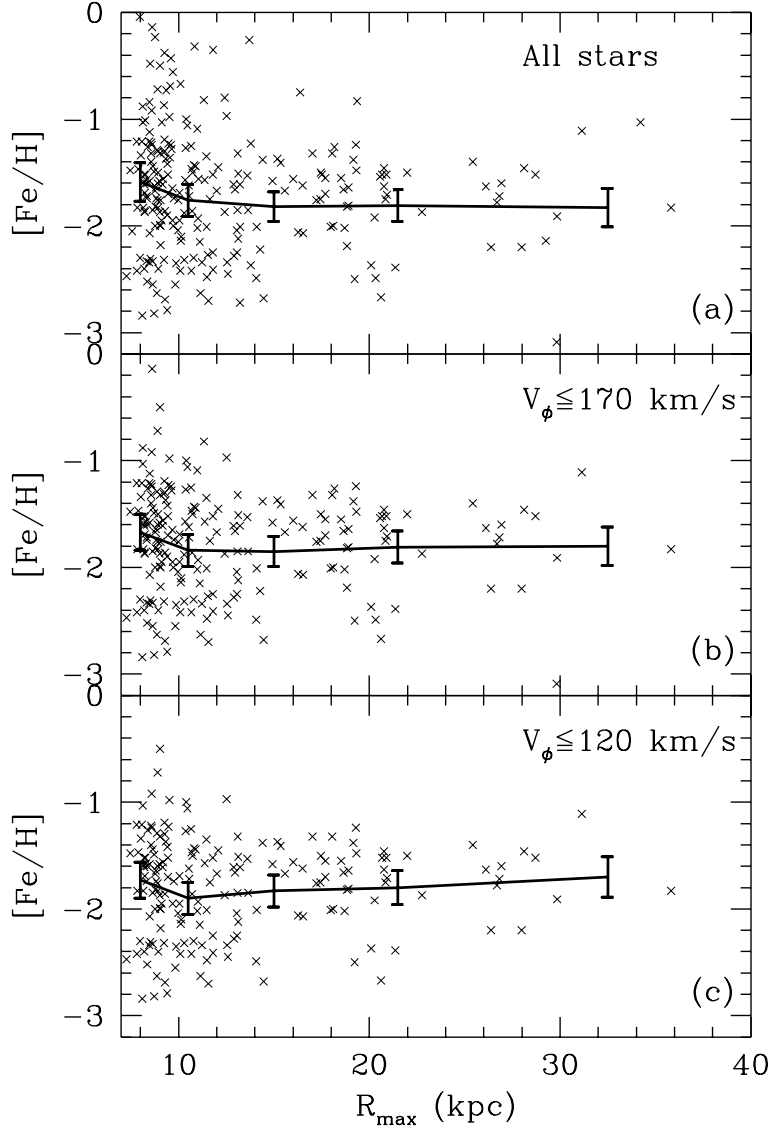}
\caption{
The relation between [Fe/H] and $R_{max}$ for the sample (crosses) with
$V_\phi\le\infty$ (a), $V_\phi\le 170$ km s$^{-1}$ (b), and $V_\phi\le 120$
km s$^{-1}$ (c). Error bars denote the mean [Fe/H] and $1\sigma$ errors
obtained in different ranges of $R_{max}$ as tabulated in Table 7, and bold
solid lines trace the mean [Fe/H]. The SLZ gravitational model is used.
}
\end{figure}

\clearpage
\begin{figure}
\plotone{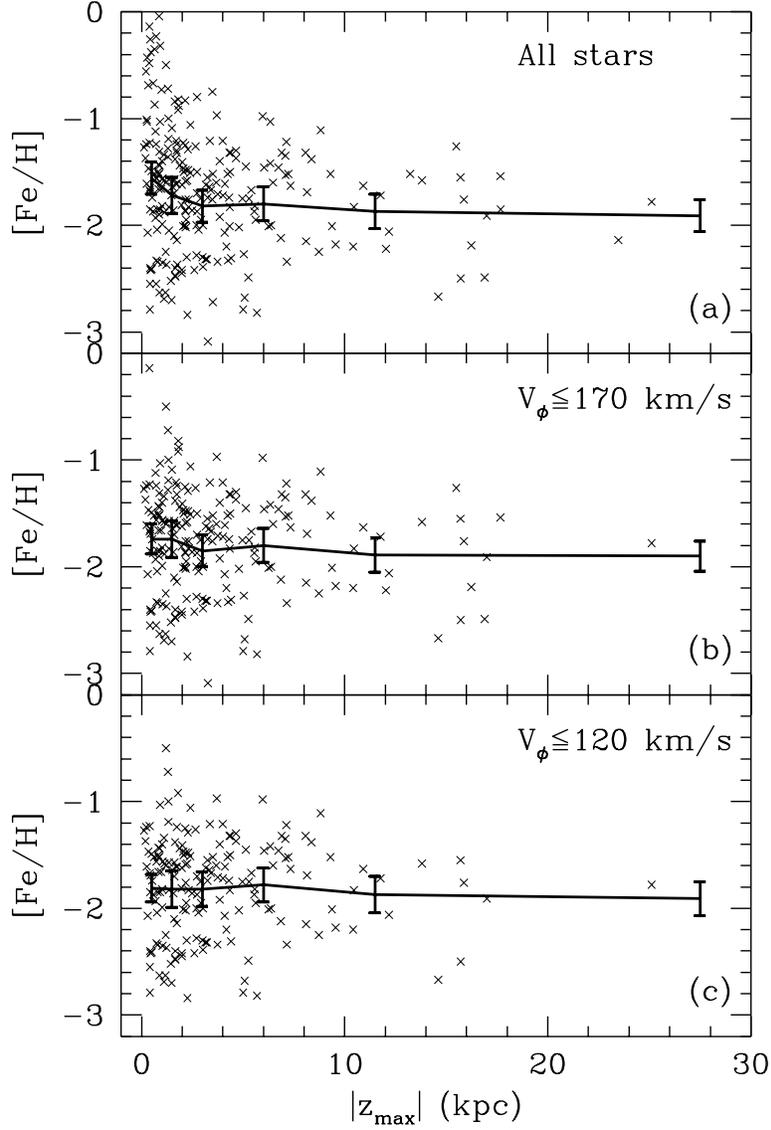}
\caption{
The relation between [Fe/H] and $|z_{max}|$ for the sample (crosses).
Error bars denote the mean [Fe/H] and $1\sigma$ errors
obtained in different ranges of $|z_{max}|$ as tabulated in Table 8.
Others are the same as for Fig. 17.
}
\end{figure}

\clearpage
\begin{figure}
\plotone{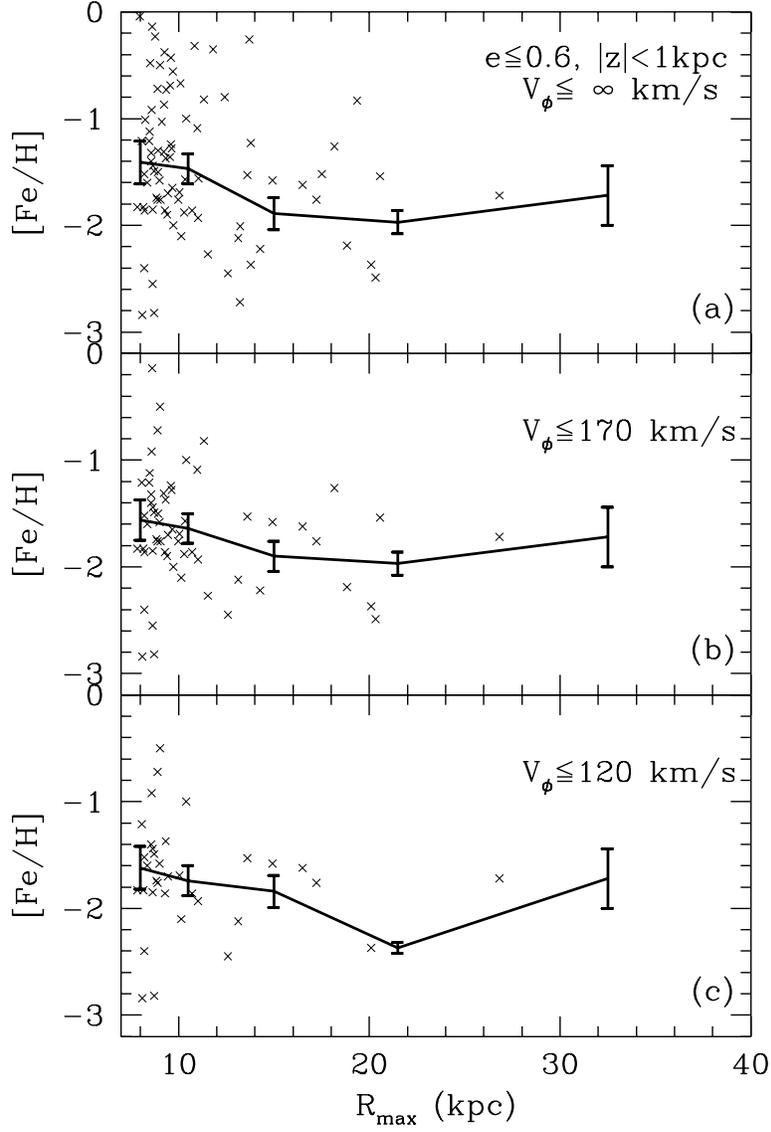}
\caption{
The same as in Fig. 17, but for the MWTD candidate stars selected from the
additional constraints of $e\le 0.6$ and $|z|<1$ kpc.
}
\end{figure}

\clearpage
\begin{figure}
\plotone{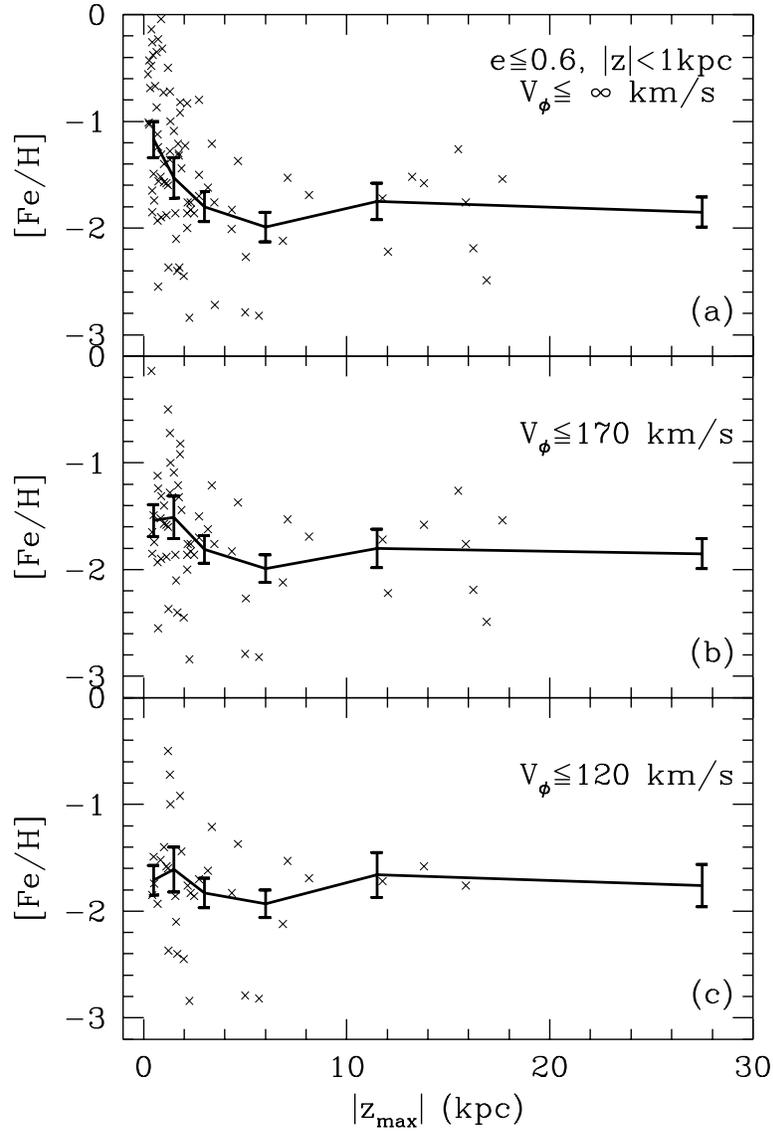}
\caption{
The same as in Fig. 19, but for [Fe/H] versus $|z_{max}|$.
}
\end{figure}

\end{document}